\documentstyle[12pt]{article}
\begin{document}

\begin{center}
{\Large
An Axiomatic Approach to Semiclassical Field Perturbation Theory
}
\\[1cm]
{{\large  O.Yu.Shvedov} \\[0.5cm]
{\it
Sub-Dept. of Quantum Statistics and Field Theory},\\
{\it Dept. of Physics, Moscow State University},\\
{\it 119992, Moscow, Vorobievy Gory, Russia}
}
\end{center}

\def\qp{
\mathrel{\mathop{\bf x}\limits^2},
\mathrel{\mathop{-i\frac{\partial}{\partial {\bf x}}}\limits^1} 
}
\def\gsim{{> \atop \sim}}
\def\lsim{{< \atop \sim}}

\setcounter{page}{0}

\begin{flushright}
hep-th/0412302
\end{flushright}

\section*{Abstract}

Semiclassical perturbation theory is investigated within the framework
of axiomatic field theory. Axioms of perturbation semiclassical theory
are formulated.  Their  correspondence with LSZ approach and Schwinger
source theory is studied.  Semiclassical S-matrix, as well as examples
of decay processes, are considered in this framework.

{\it Keywords:}
Maslov semiclassical   theory,   axiomatic   quantum   field   theory,
Bogoliubov S-matrix,  Lehmann-Symanzik-Zimmermann approach,  Schwinger
sources, Peierls brackets.

\footnotetext{e-mail:  shvedov@qs.phys.msu.su}

\footnotetext{This work was supported by the Russian  Foundation  for
Basic Research, project 02-01-01062}

\makeatletter
\@addtoreset{equation}{section}
\makeatother
\renewcommand{\theequation}{\thesection.\arabic{equation}}

\def\lb#1{\label{#1}}
\def\l#1{\lb{#1}}
\def\r#1{(\ref{#1})}
\def\c#1{\cite{#1}}
\def\i#1{\bibitem{#1}}
\def\beq{\begin{equation}}
\def\eeq{\end{equation}}
\def\bez{\begin{displaymath}}
\def\eez{\end{displaymath}}
\def\beb#1\l#2\eeb{\begin{equation} \begin{array}{c} #1 \qquad
\end{array} \label#2  \end{equation}}
\def\bey#1\eey{\begin{displaymath}
\begin{array}{c} #1  \end{array}  \end{displaymath}}

\newpage

\section{Introduction}

The main difficulty of quantum field theory (QFT) is that there is  no
nontrivial model   satisfying   all   the  axioms.  In  fact,  QFT  is
constructed within the perturbation theory framework.

However, the  perturbation  theory  is   a   partial   case   of   the
semiclassical theory.  Therefore, it is useful to generalize an axiomatic
approach to perturbation QFT to the semiclassical theory.  This is the
problem to  be  considered  in  this  paper.  The  ideas of \c{S0} are
developed here.

General structure  of  semicalssical  perturbation  theory  in  QFT is
investigated in section 2.  The main object of semiclassical theory is
{\it a semiclassical bundle} \c{S1}. Points on the space of the bundle
are interpreted  as  possible  semiclassical  states.  The base of the
bundle is the classical state space,  fibres  are  spaces  of  quantum
states in a given external classical background.

In addition  to  the "point-type" states,  one can also consider their
superpositions.

Important objects  are  introduced  in   QFT.   These   are   Poincare
transformation unitary  operators  ${\cal U}^h_g$ and Heisenberg field
operators. There analogs should arise in the semiclassical  theory  as
well. These semiclassical structures are also discussed in section 2.

Section 3  deals  with the specific features of the covariant approach
to the semiclassical field theory. Its relationship with the axiomatic
field theory \c{BLOT},  Schwinger source theory \c{Sch},  LSZ approach
\c{LSZ}, S-matrix Bogoliubov theory \c{BS,BMP}.

Section 4 is devoted to the leading order of the semiclassical theory.
All axioms of the semiclassical field theory are checked.

The semiclassical  perturbation theory is discussed in section 5.  The
calculations can be simplified, provided that the asymptotic condition
of the S-matrix approach \c{BLOT,BS,BMP,Z} is satisfied.

It is  well-known that there are difficulties of the S-matrix approach
due to unstable particles and bound states \c{BLOT}.  In section 6  we
show how  one  can  develop the semiclassical perturbation theory with
unstable particles. An example of particle decay is considered.

Section 7 contains the concluding remarks.

\section{General structure of semiclassical perturbation theory in QFT}

{\bf 2.1.}
We consider the quantum field system with the Lagrangian that  depends
on the small parameter $h$ as follows:
\beq
{\cal L} = \frac{1}{2} \partial_{\mu} \varphi \partial^{\mu} \varphi -
\frac{1}{h} V(\sqrt{h} \varphi).
\l{2.1}
\eeq
Different methods  to  develop  the semiclassical approximation theory
for this model are known.  One can use both Hamiltonian and manifestly
covariant approaches. One considers "semiclassical"
states that depend on $h$ as $h\to 0$
as follows (see \c{MS1}, \c{S2} and references therein).
For the Hamiltonian approach,
\beq
\Psi \simeq e^{\frac{i}{h}S}
e^{\frac{i}{\sqrt{h}} \int d{\bf x}
[\Pi({\bf x})  \hat{\varphi}({\bf  x})  - \Phi({\bf x}) \hat{\pi}({\bf
x})]} f \equiv K^h_{S,\Pi,\Phi} f,
\l{2.2}
\eeq
Here $\hat{\varphi}$ and $\hat{\pi}$ are field and momenta operators,
$f$ is a state vector that expands into a series in
$\sqrt{h}$. For the manifestly covariant approach,
\beq
\Psi \simeq
e^{\frac{i}{h}\overline{S}}
Te^{\frac{i}{\sqrt{h}} \int dx J(x) \hat{\varphi}(x)} \overline{f}
\equiv
e^{\frac{i}{h}\overline{S}}
T_J^h \overline{f} \equiv K^h_{\overline{S},J} \overline{f}.
\l{2.3}
\eeq
Here
$\hat{\varphi}(x)$ is a Heisenberg field,
$J(x)$ is a classical source with a compact source,  $\overline{f}$ is
expanded in $\sqrt{h}$.  A semiclassical state of the form \r{2.2} (or
\r{2.3})  can  be  viewed as a point on the space of the semiclassical
bundle \c{S1}.  Base of the bundle is set $\{ X = (S,\Pi({\bf x}),\Phi({\bf
x}))\}$  (or  $\{\overline{X}  =  (\overline{S},J(x))\}$) of classical
states,  fibers are spaces  $\{f\}$  ($\{\overline{f}\}$)  of  quantum
states in  the external field.  States \r{2.2} and \r{2.3} are written
as $K^h_X f$; the operator $K^h_X$ is called as a canonical operator.

The Maslov theory of Lagrangian manifolds  with  complex  germs
\c{M1,M2} is  a
generalization of the Maslov complex germ theory. One considers states
of the more complicated form:
\beq
\int d\alpha     K^h_{X(\alpha)}    f(\alpha),    \quad    \alpha    =
(\alpha_1,...,\alpha_k),
\l{2.4}
\eeq
They can  be   interpreted   as   $k$-dimensional   surface   on   the
semiclassical bundle.

Within the  semiclassical  theory,  one  can  formulate  the following
problems:

- let ${\cal U}_g^h$ be a Poincare transformation corresponding to  an
element $g$  of  the  Poincare  group  $G$,  $\hat{\varphi}(x)$  be  a
Heisenberg field; one should investigate these operators as $h\to 0$;

- one should find norm of the state \r{2.4} as $h\to 0$;

- one should investigate whether the expressions \r{2.3} corresponding
to different sources $J$ may approximately
coincide as $h\to 0$.

{\bf 2.2.}
It happens that the following commutation rules are satisfied:
\beb
{\cal U}^h_g K_X^h f = K^h_{u_gX} \underline{U}_g(u_gX \gets X) f;\\
\sqrt{h} \hat{\varphi}(x) K_X^h f = K_X^h \underline{\Phi}(x|X) f,
\l{2.5}
\eeb
Here $\underline{U}_g(u_gX\gets    X)$
is an unitary operator which is expanded into an asymptotic series in
$\sqrt{h}$,  $u_gX$ is a Poincare transformation of classical state,
$\underline{\Phi}(x|X)$ is an operator-valued distribution. It is also
presented as an asymptotic series in $\sqrt{h}$.  Its leading order is
a c-number quantity $\Phi(x|X)$:
\beq
\underline{\Phi}(x|X) = \Phi(x|X) + \sqrt{h} \Phi^{(1)}(x|X) + ...
\l{2.6}
\eeq

Investigate the  properties  of  the  introduced  objects.  Since  the
operators ${\cal U}_g^h$ should obey the group property
\bez
{\cal U}_{g_1g_2}^h = {\cal U}_{g_1} {\cal U}_{g_2},
\eez
it follows from eq.\r{2.5} that
\beb
u_{g_1g_2} = u_{g_1}u_{g_2};\\
\underline{U}_{g_1g_2}(u_{g_1g_2}X \gets X) =
\underline{U}_{g_1}(u_{g_1g_2}X \gets u_{g_2}X)
\underline{U}_{g_2}(u_{g_2}X \gets X).
\l{2.6z}
\eeb
Moreover, the Poincare covariance of the fields implies that
\beq
{\cal U}_{g^{-1}}^h     \hat{\varphi}(x)      {\cal      U}_g^h      =
\hat{\varphi}(w_gx), \qquad w_gx = \Lambda^{-1}(x-a)
\l{2.6w}
\eeq
Therefore, one finds from eq.\r{2.5} that
\beq
\underline{\Phi}(x|u_gX) \underline{U}_g(u_gX \gets X) =
\underline{U}_g(u_gX \gets X) \underline{\Phi}(w_gx|X).
\l{2.6v}
\eeq

{\bf 2.3.} Let us estimate the square of the norm  of state
\r{2.4} as $h\to 0$. The plan is as follows
(cf. \c{S3}). The integral
\beq
\int d\alpha  d\alpha'  (K^h_{X(\alpha)}  f(\alpha),  K^h_{X(\alpha')}
f(\alpha'))
\l{2.6c}
\eeq
is calculated with the help of the substitution
$\alpha' = \alpha + \sqrt{h}\beta$.
Then one performs an expansion in $\sqrt{h}$.
To do this, it is necessary to use the formula of expansion
of the vector
$K^h_{X(\alpha+\beta\sqrt{h})} f(\alpha     +    \beta\sqrt{h})$   in
$\sqrt{h}$.

One can obtain this formula from the commutation rule:
\beq
ih\frac{\partial}{\partial\alpha_a} K^h_{X(\alpha)}  = K^h_{X(\alpha)}
\underline{\omega}_{X(\alpha)} [\frac{\partial X}{\partial \alpha_a}].
\l{2.7}
\eeq
Here $\omega_X[\delta X]$ is an operator-valued 1-form.  It assins  an
operator in $\{f\}$ to each tangent vector $\delta X$ to the base.
In the leading order, the 1-form is a c-number:
\bez
\underline{\omega}_X[\delta X]  =  \omega_X[\delta   X]   +   \sqrt{h}
\omega_X^{(1)}[\delta X] + ...
\eez
Property $[ih\frac{\partial}{\partial \alpha_a},
ih\frac{\partial}{\partial \alpha_b}]  =  0$
implies the commutation relation for 1-forms:
\beq
\left[
\omega_{X(\alpha)} [\frac{\partial X}{\partial \alpha_a}];
\omega_{X(\alpha)} [\frac{\partial X}{\partial \alpha_b}]
\right] = - ih
\left[
\frac{\partial}{\partial \alpha_a}      \underline{\omega}_{X(\alpha)}
[\frac{\partial X}{\partial \alpha_b}]
-
\frac{\partial}{\partial \alpha_b}      \underline{\omega}_{X(\alpha)}
[\frac{\partial X}{\partial \alpha_a}]
\right],
\l{2.8}
\eeq
or
\bez
[\underline{\omega}_X[\delta X_1],\underline{\omega}_X[\delta X_2]] =
- ih d\underline{\omega}_X(\delta X_1,\delta X_2).
\eez

The operator
$K^h_{X(\alpha  +   \sqrt{h}\beta)}$   can be expanded then in
$\sqrt{h}$ as follows. Set
\beq
K^h_{X(\alpha  +   \sqrt{h}\beta)} = K^h_{X(\alpha)} V_h(\alpha,\beta).
\l{2.9}
\eeq
Differentiate \r{2.9} with respect to
$\beta_a$. Making use of eq.\r{2.7}, we obtain
\beq
\frac{\partial}{\partial\beta_a} V_h(\alpha,\beta) =
- \frac{i}{\sqrt{h}} V_h(\alpha,\beta)
\underline{\omega}_{X(\alpha + \beta \sqrt{h})}
[\frac{\partial X}{\partial \alpha_a}(\alpha + \beta \sqrt{h})].
\l{2.10}
\eeq
Therefore, the leading order in $h$ gives us a relation
\bez
V_h(\alpha,\beta) \sim    e^{-\frac{i}{\sqrt{h}}    \omega_{X(\alpha)}
[\frac{\partial X}{\partial \alpha_a}] \beta_a}.
\eez

The inner prooduct \r{2.6c} is taken to the form
\beq
h^{k/2} \int  d\alpha  d\beta  (f(\alpha),V_h(\alpha,\beta)f(\alpha  +
\beta\sqrt{h}));
\l{2.11}
\eeq
The integrand  is  a  rapidly  oscillating quantity.   Therefore,   the
integral will be exponentially small, except for the special case when
{\it the Maslov isotropic condition} is satisfied:
\beq
\omega_{X(\alpha)} [\frac{\partial X}{\partial \alpha_a}] = 0.
\l{2.12}
\eeq
For the case
\r{2.12}, the system \r{2.10}
can be   solved   within   the   perturbation   framework,   iff   the
self-consistent condition \r{2.8} is satisfied.
For the leading order, one has
\bez
V_h(\alpha,\beta) \simeq    e^{-i    \omega^{(1)}_{X(\alpha)}
[\frac{\partial X}{\partial \alpha_a}] \beta_a},
\eez
The inner product \r{2.11} takes the form
\bez
h^{k/2} \int  d\alpha  d\beta  (f(\alpha),
\prod_a \{2\pi    \delta(\omega^{(1)}_X[\frac{\partial     X}{\partial
\alpha_a}])\}
f(\alpha)).
\eez

Notice that eqs.\r{2.5} and \r{2.7} imply the relations
\beb
\underline{U}_g(u_gX \gets    X)
\underline{\omega}_X[\frac{\partial X}{\partial \alpha_a}] =
\underline{\omega}_{u_gX}[\frac{\partial (u_gX)}{\partial \alpha_a}]
\underline{U}_g(u_gX \gets X) +
ih \frac{\partial}{\partial \alpha_a} \underline{U}_g(u_gX\gets X);\\
ih\frac{\partial}{\partial \alpha_a} \underline{\Phi}(x|X) =
\left[\underline{\Phi}(x|X);
\underline{\omega}_X[\frac{\partial X}{\partial \alpha_a}]\right].
\l{2.13}
\eeb

{\bf 2.4.} In the covariant framework, some of states
\r{2.3}  approximately coincide as $h\to 0$.
This means  that  one  sould  introduce an equivalence relation of the
base of the semiclassical bundle (on the classical state space).  Some
of classical states are equivalent. Moreover, if $X_1 \sim X_2$, then
\beq
K^h_{X_1} \overline{f}_1 \simeq K^h_{X_2} \overline{f}_2
\l{2.13a}
\eeq
iff
\bez
\overline{f}_2 = \underline{V}(X_2\gets X_1) \overline{f}_1.
\eez

Investigate the properties of the operator
$\underline{V}(X_2\gets X_1)$.
First of all, the following relation
\beq
\underline{V}(X_3\gets X_1) =
\underline{V}(X_3\gets X_2)
\underline{V}(X_2\gets X_1)
\l{2.14}
\eeq
should be satisfied. Moreover, eq.
\r{2.13a} implies that
${\cal U}_g^h K^h_{X_1} \overline{f}_1 \simeq
{\cal U}_g^h K^h_{X_2} \overline{f}_2$; therefore
\beq
\underline{V}(u_gX_2 \gets u_gX_1) \underline{U}_g(u_gX_1 \gets X_1) =
\underline{U}_g(u_gX_2 \gets X_2)
\underline{V}(X_2 \gets X_1).
\l{2.15}
\eeq
It follows from the relation
$\sqrt{h} \hat{\varphi}(x) K^h_{X_1} \overline{f}_1 \simeq
\sqrt{h} \hat{\varphi}(x) K^h_{X_2} \overline{f}_2$ that
\beq
\underline{\Phi}(x|X_2) \underline{V}(X_2 \gets X_1) =
\underline{V}(X_2 \gets X_1) \underline{\Phi}(x|X_1).
\l{2.16}
\eeq
Finally, let
$(X_i,\overline{f}_i)$  depend on $\alpha$.
Differentiate \r{2.13a} with respect to $\alpha_a$:
$ih \frac{\partial}{\partial \alpha_a} K^h_{X_1} \overline{f}_1 \simeq
ih \frac{\partial}{\partial    \alpha_a}K^h_{X_2}     \overline{f}_2$;
therefore,
\beq
\underline{V}(X_2 \gets X_1)
\underline{\omega}_{X_1} [\frac{\partial X_1}{\partial \alpha_a}] =
\underline{\omega}_{X_2} [\frac{\partial X_2}{\partial \alpha_a}]
\underline{V}(X_2 \gets X_1)
+ ih \frac{\partial}{\partial \alpha_a}
\underline{V}(X_2 \gets X_1)
\l{2.17}
\eeq
provided that $X_1(\alpha) \sim X_2(\alpha)$.

{\bf 2.5.}  Thus,  all  the  problems  of  semiclassical theory can be
solved within the perturbation framework iff one specifies:

- the Poincare transformations
$u_g$ (classical) и $\underline{U}_g(u_gX
\gets X)$ (unitary operator expanded into a formal series in
$\sqrt{h}$);

- semiclassical series in $\sqrt{h}$ for $\underline{\Phi}(x|X)$ и
$\underline{\omega}_X[\delta X]$  (these opeators are c-numbers in
the leading order);

- semiclassical    series    in    $\sqrt{h}$    for   the   operators
$\underline{V}(X_2\gets X_1)$ as $X_1 \sim X_2$  (if  the  equivalence
relation is introduced on the classical state space);

These objects   should  satisfy  the  properties  \r{2.6z},  \r{2.6v},
\r{2.8}, \r{2.13}, \r{2.14}, \r{2.15}, \r{2.16}, \r{2.17}.

Therefore, let us say that a model of semiclassical field theory is
{\it given} iff the objects
$u_g$, $\underline{U}_g(u_gX \gets X)$,
$\underline{\Phi}(x|X)$, $\underline{\omega}_X[\delta X]$,
$\underline{V}(X_2\gets X_1)$ are specified and they obey the required
properties. We  suppose  the  semiclassical  model to be well-defined,
{\it even if the corresponding exact QFT model is ill-defined}.

\section{Specific features of the covariant approach to  semiclassical
perturbation theory}

{\bf 3.1.}  Let  us  investigate  the objects arising in the covariant
approach to semiclassical field theory. First, notice that eq.
\r{2.6w} implies that
\bez
{\cal U}^h_g   T^h_J   \overline{f}   =   T^h_{u_gJ}   {\cal   U}^h_g
\overline{f},
\eez
with $u_gJ(x) = J(w_gx)$, $w_g$ being of the form \r{2.6w}. Therefore,
an  explicit  form  of  transformation  $u_g$  is known,  the property
$u_{g_1g_2} = u_{g_1}u_{g_2}$ is satisifed, while the operator
$\underline{U}_g(u_gX \gets  X)  \equiv
\underline{U}_g =  {\cal  U}^h_g$  is $X$-independent and satisfies
the group property and property of invariance of the fields.
\beq
\underline{U}_{g_1g_2} = \underline{U}_{g_1} \underline{U}_{g_2},
\quad
\underline{U}_{g^{-1}} \hat{\varphi}(x)       \underline{U}_g        =
\hat{\varphi}(w_gx).
\l{3.1a}
\eeq
The 1-form $\underline{\omega}$ can be expressed via
the LSZ $R$-functions \c{LSZ}:
\beq
R(x|J) \equiv - ih (T^h_J)^+ \frac{\delta T^h_J}{\delta J(x)}.
\l{3.1}
\eeq
Namely,
\bez
ih \delta K^h_{\overline{S},J} = K^h_{\overline{S},J}
[-\delta \overline{S} - \int dx R(x|J) \delta J(x)],
\eez
therefore,
\beq
\underline{\omega}_X[\delta X] =  -  \delta  \overline{S}  -  \int  dx
R(x|J) \delta J(x),
\l{3.2}
\eeq
Moreover, $R(x|J)$ should be expanded into a formal series in
$\sqrt{h}$
\bez
R(x|J) = \overline{\Phi}(x|J) + \sqrt{h} R^{(1)}(x|J) + ...
\eez
The c-number function
$\overline{\Phi}(x|J)$ is called as a
{\it  classical field generated by the source $J$}.

Eq.\r{3.1} implies the following properties of the Hermitian
$R$-function:

- Poincare invariance
\beq
\underline{U}_{g^{-1}} R(x|u_gJ) \underline{U}_g = R(w_gx|J);
\l{3.3a}
\eeq

- Bogoliubov causality property:
\beq
\frac{\delta R(x|J)}{\delta J(y)} = 0, \qquad y\gsim x.
\l{3.3b}
\eeq

- commutation relation
\beq
[R(x|J);R(y|J)] = -ih \left(
\frac{\delta R(x|J)}{\delta J(y)} -
\frac{\delta R(y|J)}{\delta J(x)} \right);
\l{3.3c}
\eeq

- boundary condition
\beq
R(x|J) = \hat{\varphi}(x) \sqrt{h}, \qquad x \lsim supp J.
\l{3.3d}
\eeq

The operator
$\underline{\Phi}(x|J)$  can be expressed via
the $R$-function at $x\gsim supp J$:
\beq
\underline{\Phi}(x|X) = R(x|J), \qquad x\gsim supp J.
\l{3.4}
\eeq

{\bf 3.2.} Investigate the equivalence property. Say that
$J\sim 0$ iff
\beq
T^h_J \overline{f}      \simeq      e^{\frac{i}{h}     \overline{I}_J}
\underline{W}_J \overline{f}
\l{3.5}
\eeq
for some c-number phase
$\overline{I}_J$ and operator $\underline{W}_J$
presented as a formal perturbation series in $\sqrt{h}$. The following
properties  are satisfied:

- relativistic invariance:
\beq
\underline{U}_g \underline{W}_J        \underline{U}_{g^{-1}}        =
\underline{W}_{u_gJ}, \qquad \overline{I}_{u_gJ} = \overline{I}_J;
\l{3.6}
\eeq

- unitarity
\beq
\underline{W}_J^+ =
\underline{W}_J^{-1};
\l{3.7}
\eeq

- Bogoliubov causality:
if $J+\Delta J_2 \sim 0$,  $J+\Delta
J_1 + \Delta J_2 \sim 0$ and $supp \Delta J_2  \gsim  supp  \Delta  J_1$
then the operator
$(\underline{W}_{J+\Delta J_2})^+ \underline{W}_{J+\Delta J_1
+ \Delta   J_2}$   and c-number
$-   \overline{I}_{J+\Delta   J_2}    +
\overline{I}_{J+\Delta J_1 + \Delta J_2}$
do not depend on $\Delta J_2$;

- variational property:
\beq
\delta \overline{I}_J - ih \underline{W}_J^+ \delta \underline{W}_J =
\int dx R(x|J) \delta J(x);
\l{3.8}
\eeq

- boundary condition
\beq
R(x|J) =       \underline{W}_J^+       \hat{\varphi}(x)       \sqrt{h}
\underline{W}_{J}, \quad x \gsim supp J.
\l{3.9}
\eeq

It follows from eq.\r{3.9} that $\overline{\Phi}(x|J) = 0$ as  $x\gsim
supp J$. Therefore, the classical field generated by the source $J\sim
0$ has a compact support.  The  following  requirement  allows  us  to
construct  the covariant semiclassical field theory without additional
postulating equations of motion  and  commutation  relations.  Namely,
suppose  that  {\it for any field configuration $\Phi(x)$ with compact
support there exists a source $J\sim  0$  (denoted  as  $J=J_{\Phi}  =
J(x|\Phi)$)  that  generates  the  configuration  $\Phi$:  $\Phi(x)  =
\overline{\Phi}(x|J)$. It satisfies the locality property:
$\frac{\delta J(x|\Phi)}{\delta\Phi(y)} = 0$ for $x\ne y$.
}

Eq.\r{3.8} implies in the leading order in $h$ that
the functional
\beq
I[\Phi] = \overline{I}_{J_{\Phi}} - \int dx J_{\Phi}(x) \Phi(x)
\l{3.10}
\eeq
obeys the relation
\beq
J(x) = - \frac{\delta I[\Phi]}{\delta \Phi(x)}.
\l{3.11}
\eeq
It is a classical equation of motion. The functional $I[\Phi]$
satisfying the locality property
\beq
\frac{\delta^2 I}{\delta \Phi(x) \delta \Phi(y)} = 0, \quad
x\ne y
\l{3.11a}
\eeq
will be called as a {\it classical action of the theory}.

Denote $\underline{W}[\overline{\Phi}]                          \equiv
\underline{W}_{J_{\Phi}}$. The obtained relations can be formulated as
follows:

- relativistic invariance
\beq
\underline{U}_g \underline{W}[\overline{\Phi}]  \underline{U}_{g^{-1}}
= \underline{W}[u_g\overline{\Phi}];
\l{3.12a}
\eeq

- unitarity
\beq
\underline{W}^+[\overline{\Phi}] =
(\underline{W}[\overline{\Phi}] )^{-1};
\l{3.12b}
\eeq

- Bogoliubov causality
\beq
\frac{\delta}{\delta \overline{\Phi}(y)}
\left(
\underline{W}^+[\overline{\Phi}]
\frac{\delta \underline{W}[\overline{\Phi}]}
{\delta \overline{\Phi}(x)}
\right) = 0, \quad y \gsim x;
\l{3.12c}
\eeq

- the Yang-Feldman relation:
\beq
\int dy
\frac{\delta^2 I}{\delta \overline{\Phi}(x) \delta \overline{\Phi}(y)}
[R(y|J) - \overline{\Phi}(y|J)] =
ih \underline{W}^+[\overline{\Phi}]
\frac{\delta                    \underline{W}[\overline{\Phi}]}{\delta
\overline{\Phi}(x)};
\l{3.12d}
\eeq

- the boundary condition
\beq
\underline{W}^+[\overline{\Phi}] \hat{\varphi}(x) \sqrt{h}
\underline{W}[\overline{\Phi}] = R(x|J), \quad
x \gsim supp \overline{\Phi}.
\l{3.12e}
\eeq

Differential form of the Bogoliubov causality  property  \r{3.12c}  is
obtained by a standard procedure.  The Yang-Feldman relation \r{3.12d}
is derived from the variational property \r{3.8} with the help of  the
substitution
\bez
\delta J(x) = - \int dy
\frac{\delta^2 I}{\delta \overline{\Phi}(x) \delta \overline{\Phi}(y)}
\delta \overline{\Phi}(y).
\eez

{\bf 3.3.} It happens that all  objects  of  the  semiclassical  field
theory considered  in  the previous subsection can be reconstructed if
one specifies:

- action $I[\Phi]$ satisfying the  locality  and  Poincare  invariance
property;

- operators   $\hat{\varphi}(x)$  and  $\underline{U}_g$  expanded  in
$\sqrt{h}$ and satisfying the properties \r{3.1a};

- Hermitian operators $R(x|J)$ expanded in $\sqrt{h}$  and  satisfying
the properties \r{3.3a},  \r{3.3b}, \r{3.3c}, \r{3.3d}; in the leading
order,  the  operators  $R(x|J)$  should  be  equal  to  the  solution
$\overline{\Phi}(x|J)$      of     eq.\r{3.11}     under     condition
$\overline{\Phi}|_{x\lsim supp J} = 0$;

- operator $\underline{W}[\overline{\Phi}]$ expanded in $\sqrt{h}$ and
satisfying the relations \r{3.12a},  \r{3.12b},  \r{3.12c}, \r{3.12d},
\r{3.12e}.

These properties are not independent.  In section 5, we will show that
they are  related  with  each  other.  Let  us  show  now  how one can
reconstruct all the structures of the semiclassical field  theory  and
check the  properties  of  section  2.  Let  us  consider the simplest
example - the theory with classical action
\beq
I[\Phi] = \int dx [\frac{1}{2} \partial_{\mu} \Phi \partial^{\mu} \Phi
- V(\Phi)],
\l{3.13}
\eeq
where $V(\Phi) \sim \frac{m^2}{2} \Phi^2$, $\Phi \to 0$.

Notice that  the  classical   Poincare   transformation   is   already
constructed. Namely,
$u_gJ(x) = J(w_gx)$;  $\overline{U}_g$  is $X$-independent.
The 1-form
$\underline{\omega}$ is of the form \r{3.2}.

Say that $J\sim J'$ iff for some source
$J_+$ with  the support $supp J_+ > supp J$,  $supp J_+ > supp J'$ the
properties
$J+J_+   \sim   0$,  $J'  +  J_+  \sim  0$ are satisfied.
This definition is equivalent to the following:
\beq
\overline{\Phi}(x|J) = \overline{\Phi}(x|J'), \quad
x \gsim supp J,J' \quad \Leftrightarrow \quad J \sim J'
\l{3.14}
\eeq
Say that
$(\overline{S}_1,J_1) \sim (\overline{S}_2,J_2)$ iff
\beq
\overline{S}_1 + \overline{I}_{J_1+J_+} =
\overline{S}_2 + \overline{I}_{J_2+J_+}
\l{3.14a}
\eeq
This definition does not depend on the particular choice of
$J_+$.  Consider the operator
$\underline{V}(J_2 \gets J_1)$ of the form
\beq
\underline{V}(J_2 \gets J_1) = \underline{W}^+_{J_2+J_+}
\underline{W}_{J_1 + J_+},
\l{3.15}
\eeq
Formula \r{3.15}  is  also  well-defined  due  to the
Bogoliubov causality
condition. Define the operatoe
$\underline{\Phi}(x|J)$ as follows.
If $x\gsim supp J$ then define
$\underline{\Phi}(x|J)$ by relation \r{3.4}. For the general case,
choose a source $J'\sim  J$ such that
$supp J' \lsim x$ and set
\beq
\underline{\Phi}(x|J) =     \underline{V}(J    \gets    J')    R(x|J')
\underline{V}(J' \gets J).
\l{3.16}
\eeq

One chould check:

- whether  eq.\r{3.16}  is  well-defined;  this  is a corollary of the
identity
\beq
R(x|J_2) \underline{V}(J_2 \gets J_1) = \underline{V}(J_2  \gets  J_1)
R(x|J_1), \quad x \gsim supp J_{1,2};
\l{3.17}
\eeq

- property \r{2.6z} (it is formulated as an axiom - eq.\r{3.1a});

- relation \r{2.8} (it is reduced to eq.\r{3.3c});

- eqs.\r{2.15}  (they are corolloaries of the Poincare covariance
of $\underline{W}$);

- eq.\r{2.17}, which is of the form
\beb
\underline{V}(J_2 \gets J_1)
[-\delta \overline{S}_1 - \int dx R(x|J_1) \delta J_1(x)]
\\ =
[-\delta \overline{S}_2 - \int dx R(x|J_2) \delta J_2(x)]
\underline{V}(J_2 \gets J_1)
\\
+ ih \delta \underline{V}(J_2 \gets J_1);
\l{3.18}
\eeb

- property \r{2.16} (it is a corollary of \r{3.17});

- relations \r{2.13},  the first  of  them  is  a  corollary  of  Poincare
invariance, the second one can be presented as
\beq
ih \delta  \underline{\Phi}(x|J)  =  [\underline{\Phi}(x|J);  -\int dy
R(y|J) \delta J(y)];
\l{3.19}
\eeq

- eq.\r{2.6v} (it is a corollary of Poincare invariance of
$R$ and $W$).

Thus, to check the axioms of section 2, one should justify relations
\r{3.17}, \r{3.18}, \r{3.19}.

Eq.\r{3.18} is taken to the form
\beb
\delta \overline{S}_1 + \int dx \overline{\Phi}(x|J_1) \delta J_1(x) =
\delta \overline{S}_2 + \int dx \overline{\Phi}(x|J_2) \delta J_2(x);\\
\quad \\
-ih \delta \underline{W}_{J_2+J_+} \underline{W}^+_{J_2+J_+}
+ ih \delta \underline{W}_{J_1+J_+} \underline{W}^+_{J_1+J_+}
=\\
- \underline{W}_{J_1+J_+}
\int dx (R(x|J_1) - \overline{\Phi}(x|J_1)) \delta J_1(x)
\underline{W}_{J_1+J_+}^+
\\+
\underline{W}_{J_2+J_+}
\int dx (R(x|J_2) - \overline{\Phi}(x|J_2)) \delta J_2(x)
\underline{W}_{J_2+J_+}^+
\l{3.20}
\eeb
here $J_+$  and  $J_+  +  \delta  J_+$  are sources found from the
relations $J_1+J_+ \sim 0$, $J_2+J_+\sim 0$, $J_1 + \delta J_{1,2} +
J_+ + \delta J_+ \sim 0$; the supports of the
sources satisfy the conditions
\bez
supp \delta  J_+ \gsim supp J_+ \gsim supp J \cup supp \delta J_1 \cup
supp \delta J_2.
\eez
The Yang-Feldman relation \r{3.12d} implies that
\beq
-ih \underline{W}^+_J \delta \underline{W}_J =
\int dx [R(x|J) - \overline{\Phi}(x|J)] \delta J(x)
\eeq
for $J  \sim  0$,  $J+\delta  J  \sim 0$.  Therefore, relation
\r{3.20} is equivalent to the following one:
\bey
0 =
- \underline{W}_{J_1+J_+}
\int dx (R(x|J_1+J_+) - \overline{\Phi}(x|J_1+J_+)) \delta J_+(x)
\underline{W}_{J_1+J_+}^+
\\
+ \underline{W}_{J_2+J_+}
\int dx (R(x|J_2+J_+) - \overline{\Phi}(x|J_2+J_+)) \delta J_+(x)
\underline{W}_{J_2+J_+}^+.
\eey
This property is valid  because  of  conditions  on  the  support  and
boundary  condition \r{3.12e}.  Thus,  the second relation \r{3.20} is
checked, while the first one is a corollary of eq. \r{3.14a}.

Property \r{3.17} is a corollary of \r{3.18}. Namely, one can choose a
source $\delta J$ such that $supp \delta J \gsim supp J_{1,2}$.

Let us justify property \r{3.19}.  Consider the partial case
$x\gsim supp J$, $x\gsim supp \delta J$. Then relation
\beq
ih\delta R(x|J) =
[R(x|J), -\int dy R(y|J) \delta J(y)], \quad x \gsim supp J,\delta J
\l{3.22}
\eeq
is a corollary of the commutation  relation  \r{3.3c}  and  Bogoliubov
causality property \r{3.3b},

Consider now the general case. Choose sources
$J'$ and $\delta J'$ such that
$J' \sim J$, $J'+\delta J' \sim J + \delta J$,
$supp J' \lsim x$, $supp \delta J' \lsim x$. Set
\bez
\underline{\Phi}(x|J) = \underline{V}(J\gets J') R(x|J')
\underline{V}(J' \gets J).
\eez
Then eqs. \r{3.18} and \r{3.22} imply property \r{3.19}.

\section{The leading order of semiclassical expansion}

{\bf 4.1.}
In the previous section we have obtained the following  properties  of
the operators:
\bey
\hat{\varphi}(x) \simeq \hat{\varphi}_0(x),
\quad \underline{U}_g \simeq U_g, \\
\underline{W}[\overline{\Phi}] \simeq
{W}[\overline{\Phi}], \quad
R(x|J) \simeq \overline{\Phi}(x|J) + \sqrt{h} R^{(1)}(x|J)
\eey
Let us write them in the leading order of the semiclassical expansion.
For $R^{(1)}$, one has:
\beb
U_{g_1g_2} = U_{g_1}U_{g_2},\quad
U_{g^{-1}} R^{(1)} (x|u_gJ) U_g = R^{(1)}(w_gx|J);
\\ \qquad \\
\left[ R^{(1)}(x|J);R^{(1)}(y|J) \right]
= - i(D_{\overline{\Phi}_J}^{ret}(x,y) -
D_{\overline{\Phi}_J}^{ret}(y,x)) \equiv
-iD_{\overline{\Phi}_J}(x,y); \\
R^{(1)}(x|J) = \hat{\varphi}_0(x), \quad x\lsim supp J;\\
\frac{\delta R^{(1)}(x|J)}{\delta J(y)} = 0, \quad y \gsim x;
\l{4.1}
\eeb
Here $D^{ret}_{\overline{\Phi}}$ is a retarded Green function
for the equation
\beq
(\partial_{\mu} \partial^{\mu}   +  V'{}'(\overline{\Phi}(x)))  \delta
\Phi(x) = \delta J(x).
\l{4.2}
\eeq
For $W[\overline{\Phi}]$, one has:
\beb
U_g W[\overline{\Phi}] U_{g^{-1}} = W[u_g\overline{\Phi}],
\quad
W^+[\overline{\Phi}] = (W[\overline{\Phi}])^{-1};\\
\frac{\delta}{\delta \overline{\Phi}(y)}
\left(
W^+[\overline{\Phi}] \frac{\delta           W[\overline{\Phi}]}{\delta
\overline{\Phi}(x)}
\right) = 0, \quad y \gsim x.
\l{4.3}
\eeb
For the case $J\sim 0$, the operator
$R^{(1)}(x|J)$ should satisfy the following
equations:
\beq
(\partial_{\mu} \partial^{\mu}      +     V'{}'(\overline{\Phi}_J(x)))
R^{(1)}(x|J) = 0; \\
R^{(1)}(x|J) =         W^+[\overline{\Phi}]         \hat{\varphi}_0(x)
W[\overline{\Phi}], \quad x \gsim supp \overline{\Phi}.
\l{4.4}
\eeq
Let us construct  the  objects  obeying  relations  \r{4.1},  \r{4.3},
\r{4.4}.  Because  of  the boundary conditions on $R^{(1)}$ at $x\lsim
supp J$,  the field $\hat{\varphi}_0(x)$ satisfies  the  equations  of
motion  and  commutation relations for the free theory with the square
mass $m^2 = V'{}'(0)$. When one chooses the Fock representation of the
canonical commutation relations, the oparator
$\hat{\varphi}_0(x)$ coincides with the free scalar field with the mass
$m$, while the operator $U_g$ is a Poincare transformation in the free
theory.

By
$\hat{\phi}_{\pm}(x|\overline{\Phi})$
we denote the solution of the problem
\beb
(\partial_{\mu} \partial^{\mu}      +     V'{}'(\overline{\Phi}(x)))
\hat{\phi}_{\pm}(x|\overline{\Phi}) = 0,
\\
\hat{\phi}_-|_{x\lsim supp \overline{\Phi}} = \hat{\varphi}_0,
\qquad
\hat{\phi}_+|_{x\gsim supp \overline{\Phi}} = \hat{\varphi}_0.
\l{4.5}
\eeb
Then
\bez
R^{(1)}(x|J) = \hat{\phi}_{-}(x|\overline{\Phi}_J).
\eez
The properties of  Poincare  invariance  and  causality  are  evident;
commutation relations
\r{4.1} are checked as follows:  one considers the difference  between
sides of equation
$\Delta_{\overline{\Phi}}(x,y)$  and obtains the set of equations:
\bey
(\partial_{\mu}^x \partial^{\mu}_x      +     V'{}'(\overline{\Phi}(x)))
\Delta_{\overline{\Phi}}(x,y) = 0,
\\
(\partial_{\mu}^y \partial^{\mu}_y      +     V'{}'(\overline{\Phi}(y)))
\Delta_{\overline{\Phi}}(x,y) = 0,\\
\Delta_{\overline{\Phi}}(x,y) =    0,\qquad     x,y     \lsim     supp
\overline{\Phi}.
\eey
One then finds that $\Delta_{\overline{\Phi}}(x,y) = 0$.

The boundary condition \r{4.4} can be also presented as
\beq
\hat{\phi}_{+}(x|\overline{\Phi}) W[\overline{\Phi}]
= W[\overline{\Phi}] \hat{\phi}_{-}(x|\overline{\Phi})
\l{4.6}
\eeq

Consider whether there exists an unitary operator $W[\overline{\Phi}]$
satisfying the set of equations \r{4.3},  \r{4.6}.  First, notice that
it     is     defined     up    to    a    c-number    phase    factor
$e^{i\gamma[\overline{\Phi}]}$.   The   phase   should   be   Poincare
invariant. Moreover, the causality property \r{4.3} imply the locality
property                                               $\frac{\delta^2
\gamma[\overline{\Phi}]}{\delta\overline{\Phi}(x)               \delta
\overline{\Phi}(y)} = 0,  \quad x\ne y$.  This phase factor is related
to one-loop renormalization.

{\bf 4.2.} Investigate whether there exists an unitary operator
$W[\overline{\Phi}]$ satisfying relation \r{4.6}. The operators
$\hat{\phi}_{+}(x|\overline{\Phi})$
and $\hat{\phi}_{-}(x|\overline{\Phi})$
obey the same commutation relation
\bez
[\hat{\phi}_{\pm}(x|\overline{\Phi});
\hat{\phi}_{\pm}(x|\overline{\Phi})] =                     \frac{1}{i}
D_{\overline{\Phi}}(x,y),
\eez
It is checked analogously to
\r{4.1}. Therefore, the correspondence between operators
$\hat{\phi}_{\pm}$ is a linear canonical
transformation. To check that  this  transformation  is  unitary,  one
should justify the Berezin condition \c{Ber}.

Denote
\beq
v(x) = V'{}'(\overline{\Phi}(x)) - m^2;
\l{4.6a}
\eeq
then
\bey
\hat{\phi}_{+}(x|\overline{\Phi}) = \hat{\varphi}_0(x)
- \int dy D_{\overline{\Phi}}^{adv}(x,y) v(y) \hat{\varphi}_0(y),
\\
\hat{\phi}_{-}(x|\overline{\Phi}) = \hat{\varphi}_0(x)
- \int dy D_{\overline{\Phi}}^{ret}(x,y) v(y) \hat{\varphi}_0(y),
\eey
where $D_{\overline{\Phi}}^{ret}(x,y)$ and
$D_{\overline{\Phi}}^{adv}(x,y)$ are advanced and
retarded Green functions for eq \r{4.2}. For the case
$x\gsim supp v$, relation \r{4.6} can be written as
\beq
W^+[\overline{\Phi}] \hat{\varphi}_0(x) W[\overline{\Phi}] =
\hat{\varphi}_0(x)
- \int dy D_{\overline{\Phi}}^{ret}(x,y) v(y) \hat{\varphi}_0(y),
\quad x \gsim supp v.
\l{4.7}
\eeq
The free-field  creation  and  annihilation operators can be expressed
via the field operators as
\beq
a_{{\bf p}}^{\pm} = \int_{x^0=const} d{\bf x}
\frac{1}{(2\pi)^{d/2}}
\sqrt{\frac{\omega_{{\bf p}}}{2}}
\left\{
\hat{\varphi}_0(x) \mp             \frac{i}{\omega_{\bf            p}}
\frac{\partial}{\partial x^0} \hat{\varphi}_0(x)
\right\} e^{\mp ipx},
\l{4.8}
\eeq
Therefore, eq.\r{4.7} is equivalent to the following relation:
\bey
W^+[\overline{\Phi}] a^{\pm}_{\bf p} W[\overline{\Phi}] =
a^{\pm}_{\bf p} -
\frac{1}{(2\pi)^{d/2}} \sqrt{\frac{\omega_{\bf p}}{2}}
\int_{x^0=const} d{\bf x} dy
v(y) \hat{\varphi}_0(y) e^{\mp ipx}
\left\{
1 \mp \frac{i}{\omega_{\bf p}} \frac{\partial}{\partial x^0}
\right\} D^{ret}_{\overline{\Phi}}(x,y).
\eey
Therefore,
\beq
W^+[\overline{\Phi}] a^{\pm}_{\bf p} W[\overline{\Phi}] =
a^{\pm}_{\bf p} + \int d{\bf k}
[A^{\pm}_{{\bf pk}}    a^{\pm}_{{\bf    k}}   +   B^{\pm}_{{\bf   pk}}
a^{\mp}_{\bf k}],
\l{4.9}
\eeq
where the c-number functions $A^{\pm}_{{\bf pk}}$ and
$B^{\pm}_{{\bf pk}}$ rapidly damps
at ${\bf  p},{\bf  k}  \to\infty$
as Fourier   transformations   of   functions  with  compact  support.
According to Berezin condition, there exists
$W[\overline{\Phi}]$ satisfying eq.\r{4.9} iff
$B^{\pm}_{{\bf pk}} \in L^2$.  For our case, this condition
is satisfied; therefore, the operator $W$ obeying eq.\r{4.6} exists.

Write
\beq
W[\overline{\Phi}] = c[\overline{\Phi}] W^0[\overline{\Phi}],
\l{4.10}
\eeq
where $c[\overline{\Phi}]$ is a c-number factor;  it absolute value is
fixed due to unitarity, while
$W^0[\overline{\Phi}]$ is an operator determined uniquely from eq.
\r{4.6}  and normalization condition
\beq
<0|W^0[\overline{\Phi}]|0> = 1.
\l{4.11}
\eeq
Investigate the properties \r{4.3}.  The Poincare invariance
relation can be rewritten as
\beq
c[u_g\overline{\Phi}] = c[\overline{\Phi}],
\l{4.12}
\eeq
since the operator $U_{g^{-1}}W[u_g\overline{\Phi}]U_g$  satisfis  the
commutation  relations \r{4.6} and coincides with $W[\overline{\Phi}]$
up to a c-multiplier.

To investigate  unitarity  and  causality  properties,  introduce  the
current operator
\beq
\hat{\rho}(x|\overline{\Phi}) \equiv
\frac{1}{i} W^+[\overline{\Phi}] \frac{\delta
W[\overline{\Phi}]}{\delta \overline{\Phi}(x)}.
\l{4.13}
\eeq
If $W$ is an unitary operator then the current is Hermitian.  When one
mmultiplies  the  operator  $W[\overline{\Phi}]$ by a c-number factor,
the  operator  $\hat{\rho}(x|\overline{\Phi})$  is  increased   by   a
c-number; therefore, the multiplier $c[\overline{\Phi}]$ satisfies the
unitarity condition iff
\beq
<0|\hat{\rho}(x|\overline{\Phi}|0> =
<0|\hat{\rho}^+(x|\overline{\Phi}|0>.
\l{4.14}
\eeq
The causality condition means that
\beq
\frac{\delta \hat{\rho}(x|\overline{\Phi})}{\delta \overline{\Phi}(y)}
= 0, \qquad y \gsim x.
\l{4.15}
\eeq

{\bf 4.3.}   Let   us   obtain   an  explicit  form  of  the  operator
$\hat{\rho}(x|\overline{\Phi})$. Differentiating relation \r{4.6} with
respect to $\overline{\Phi}$, we obtain that
\beb
W^+[\overline{\Phi}]
\frac{\delta
\hat{\phi}_+(x|\overline{\Phi})
}{\delta
\overline{\Phi}(y)} W[\overline{\Phi}] +
i \hat{\phi}_-(x|\overline{\Phi})
\hat{\rho}(y|\overline{\Phi}) =
i \hat{\rho}(y|\overline{\Phi})
\hat{\phi}_-(x|\overline{\Phi}) +
\frac{\delta
\hat{\phi}_-(x|\overline{\Phi})
}{\delta
\overline{\Phi}(y)}.
\l{4.16}
\eeb
Variating eq.\r{4.5}, we find:
\beb
\delta \hat{\phi}_-(x|\overline{\Phi}) =
- \int D^{ret}_{\overline{\Phi}}(x,y)
V'{}'{}' (\overline{\Phi}(y))
\hat{\phi}_-(y|\overline{\Phi})
\delta \overline{\Phi}(y) dy;
\\
\delta \hat{\phi}_+(x|\overline{\Phi}) =
- \int D^{adv}_{\overline{\Phi}}(x,y)
V'{}'{}' (\overline{\Phi}(y))
\hat{\phi}_+(y|\overline{\Phi})
\delta \overline{\Phi}(y) dy
\l{4.17}
\eeb
Therefore,
\beq
i [\hat{\rho}(y|\overline{\Phi});  \hat{\phi}_-(x|\overline{\Phi})]  =
D_{\overline{\Phi}}(x,y)
V'{}'{}'(\overline{\Phi}(y))
\hat{\phi}_-(y|\overline{\Phi}).
- \int D^{ret}_{\overline{\Phi}}(x,y)
\l{4.18}
\eeq
Thus, the operator
$\hat{\rho}(y|\overline{\Phi})$
is reconstructed up to an additive constant.  It can be expressed  via
the normal    products \c{BS,Z}.
As   usual,   define   the   positive-   and
negative-frequency
(at $-\infty$)
parts
$\hat{\phi}_-^{\pm}(y|\overline{\Phi})$ of the field
$\hat{\phi}_-(y|\overline{\Phi})$ due to the relations
\beq
(\partial_{\mu} \partial^{\mu}      +     V'{}'(\overline{\Phi}(x)))
\hat{\phi}_-^{\pm}(x|\overline{\Phi}) = 0,
\quad
\hat{\phi}_-^{\pm}|_{x\lsim supp          \overline{\Phi}}           =
\hat{\varphi}_0^{\pm}.
\l{4.19}
\eeq
Therefore,
\beq
\hat{\rho}(y|\overline{\Phi}) =             -
- \frac{1}{2} V'{}'{}'(\overline{\Phi}(y))
\left[
:\hat{\phi}_-^2(y|\overline{\Phi}): + \alpha(y|\overline{\Phi})
\right],
\l{4.20}
\eeq
where $\alpha(y|\overline{\Phi})$ is a
multiplicator by a real number.
Consider the vacuum
averaging values of left-hand and right-hand sides of the
relations
\bez
iW[\overline{\Phi}] \hat{\rho}(x|\overline{\Phi}) =
\frac{\delta W[\overline{\Phi}]}{\delta \overline{\Phi}(x)}.
\eez
We obtain the relation for finding
$c[\overline{\Phi}]$:
\beb
\frac{\delta c[\overline{\Phi}]}{\delta \overline{\Phi}(x)} =
i <0| W[\overline{\Phi}] \hat{\rho}(x|\overline{\Phi})|0> =
\\ - \frac{i}{2} V'{}'{}'(\overline{\Phi}(x))
<0|W[\overline{\Phi}]
\left[
\hat{\phi}_-^+(x|\overline{\Phi}) \hat{\phi}_-^+(x|\overline{\Phi})
+ \alpha(y|\overline{\Phi})
\right]|0>.
\l{4.21}
\eeb
Investigate the     commutation      rules      between      operators
$W[\overline{\Phi}]$ and $\hat{\phi}_-^{\pm}$. First, write
\bez
\hat{\phi}_-(x|\overline{\Phi}) = \hat{\varphi}_0(x) -
\int dy D_0^{ret}(x,y) v(y) \hat{\phi}_-(y|\overline{\Phi});
\eez
therefore,
for $x \gsim supp v$ one finds from \r{4.6} that
\beq
W^+[\overline{\Phi}] \hat{\varphi}_0(x) W[\overline{\Phi}] =
\hat{\varphi}_0(x)
- \int dy D_0^{ret}(x,y) v(y) \hat{\phi}_-(y|\overline{\Phi}),
\quad x \gsim supp v.
\l{4.22}
\eeq
Notice that    the    positive-frequency    part   of   the   function
$D_0^{ret}(x,y)$ coincides  with  $D_0^+(x,y)$  at  $x\gsim  supp  v$.
Therefore, it follows from eq.\r{4.22} that
\beb
W^+[\overline{\Phi}] \hat{\varphi}_0^+(x) W[\overline{\Phi}] =
\hat{\varphi}_0^+(x)
- \int dy D_0^{+}(x,y) v(y) \hat{\phi}_-(y|\overline{\Phi}) \\
=
\hat{\phi}_-^+(x|\overline{\Phi}) +
\int dy        (D_0^{ret}(x,y)        -        D_0^+(x,y))        v(y)
\hat{\phi}_-^+(y|\overline{\Phi}) -
\int dy D_0^+(x,y) v(y) \hat{\phi}_-^-(y|\overline{\Phi}).
\\ \quad
\l{4.23}
\eeb
Notice also that
\bez
D_0^{ret} - D_0^+ = D_0^c.
\eez

Investigate the corrolaries of relation \r{4.23}.

Denote by $\check{A}$ the operator with the kernel $A(x,y)$.  Multiply
eq.\r{4.23} by
$W[\overline{\Phi}]$  and use relation
$<0|\hat{\varphi}_0^+(x) = 0$. We obtain that
\bez
0 = <0|W[\overline{\Phi}] \hat{\phi}_-^+(x|\overline{\Phi}) -
<0|W[\overline{\Phi}]
\left\{
(\check{1} +  \check{D}_0^c  \check{v})^{-1}  \check{D}_0^+  \check{v}
\check{\phi}_-
\right\}(x|\overline{\Phi})
\eez
Here $\check{v}$ is the operator of multiplication by
$v(x)$ with the kernel
$v(x,y) = v(x) \delta(x-y)$.

Relation \r{4.21} is taken to the form
\beq
\frac{\delta c[\overline{\Phi}]}{\delta \overline{\Phi}(x)} =
- \frac{i}{2} c[\overline{\Phi}] V'{}'{}'(\overline{\Phi}(x))
\left[
\frac{1}{i}
\{
(\check{1} +  \check{D}_0^c  \check{v})^{-1}  \check{D}_0^+  \check{v}
\check{D}^-_{\overline{\Phi}}
\}(x,x) + \alpha(x|\overline{\Phi})
\right].
\l{4.24}
\eeq

Here $D_{\overline{\Phi}}^-$ is a commutation function
for the operators
$\hat{\phi}_-^-$ и $\hat{\phi}_-^+$:
\bez
[\hat{\phi}^-_-(\xi_1|\overline{\Phi});
\hat{\phi}^+_-(\xi_2|\overline{\Phi})] =
\frac{1}{i}
D_{\overline{\Phi}}^-(\xi_1,\xi_2).
\eez
Eq.\r{4.24} can be also rewritten as
\beq
\alpha\{x|v\} = 2i \frac{\delta \ln c[v]}{\delta v(x)} +
i
\{
(\check{1} +  \check{D}_0^c  \check{v})^{-1}  \check{D}_0^+  \check{v}
\check{D}^-_{\overline{\Phi}}
\}(x,x).
\l{4.25}
\eeq
The following identity
\bez
(\check{1} +  \check{D}_0^c  \check{v})^{-1}  \check{D}_0^+  \check{v}
\check{D}^-_{\overline{\Phi}}
=
- \check{D}^-_{\overline{\Phi}} +
(\check{1} +  \check{D}_0^c  \check{v})^{-1}  \check{D}_0^c
-
(\check{1} +  \check{D}_0^{adv}  \check{v})^{-1}  \check{D}_0^{adv}.
\eez
is checked by a  direct  calculation.  The  last  term  satisfies  the
property
$\{(\check{1} +            \check{D}_0^{adv}            \check{v})^{-1}
\check{D}_0^{adv}\}(x,x) = 0$;
therefore,
\beq
\alpha\{x|v\} = 2i \frac{\delta \ln c[v]}{\delta v(x)} +
i
\{
- \check{D}^-_{\overline{\Phi}}
+
(\check{1} +  \check{D}_0^c  \check{v})^{-1}  \check{D}_0^c
\}(x,x).
\l{4.26}
\eeq
Since $Re {D}_{\Phi}^-(x,x) = 0$,
$\frac{\delta {D}_{\Phi}^-(x,x)}{\delta v(y)} =  0$  at  $y\gsim  x$,
the Poincare  invariance,  unitarity  and  causality properties can be
presented as
\beb
c[u_gv] = c[v];\\
Re \left\{
\frac{\delta c[v]}{\delta v(x)} +
\frac{1}{2}
[(\check{1} +  \check{D}_0^c  \check{v})^{-1}  \check{D}_0^c](x,x)
\right\} = 0; \\
\frac{\delta}{\delta v(y)}
\left\{
\frac{\delta c[v]}{\delta v(x)} +
\frac{1}{2}
[(\check{1} +  \check{D}_0^c  \check{v})^{-1}  \check{D}_0^c](x,x)
\right\} = 0, \quad y \gsim x.
\l{4.27}
\eeb
Formally, the solution of system \r{4.27} has the form
\bez
c[v] = (det (\check{1} +  \check{D}_0^c  \check{v}))^{-1/2}
e^{i\gamma[v]}.
\eez
Here
$\gamma[v]$  is an arbitrary real local functional satisfying the
condition
$\frac{\delta^2\gamma[v]}{\delta  v(x)\delta
v(y)} = 0$ as $x\ne y$.
Relation \r{4.27} may be viewed as a definition of the
determinant. Although the quantity
$[(\check{1} +   \check{D}_0^c   \check{v})^{-1}  \check{D}_0^c](x,x)$
diverges, its real part and variation derivative with respect to
$v(y)$ at $y\ne x$ are finite.

Eq.\r{4.25} implies the important property:
\beb
\frac{\delta \alpha\{x|v\}}{\delta v(y)} -
\frac{\delta \alpha\{y|v\}}{\delta v(x)} =
-i \left(
\frac{\delta D_{\overline{\Phi}}^-(x,x)}{\delta v(y)} -
\frac{\delta D_{\overline{\Phi}}^-(y,y)}{\delta v(x)}
\right) =
i [(D_{\overline{\Phi}}^-(x,y))^2 - (D_{\overline{\Phi}}^-(y,x))^2]
\\ \quad
\l{4.28}
\eeb
This will be used further.

Relation \r{4.28} can be also obtained from the identity
\beq
[\hat{\rho}(x|\overline{\Phi};\hat{\rho}(y|\overline{\Phi}] =
- i
\left(
\frac{\delta \hat{\rho}(x|\overline{\Phi})}{\delta \overline{\Phi}(y)}
-
\frac{\delta \hat{\rho}(y|\overline{\Phi})}{\delta \overline{\Phi}(x)}
\right).
\l{4.29a}
\eeq
It follows from eq.\r{4.28} and causality condition that
\beq
\frac{\delta \alpha\{x|v\}}{\delta v(y)} =
i [(D_{\overline{\Phi}}^-(x,y))^2 - (D_{\overline{\Phi}}^-(y,x))^2]
\theta(x^0-y^0).
\l{4.29}
\eeq
For $x=y$, one should clarify relation
\r{4.29}. This depends on the choice of one-loop counterterm.

\section{Semiclassical $S$-matrix}

{\bf 5.1.} Let us analyze the semiclassical perturbation theory. It is
necessary to fix the representation.  Otherwise,  the  next  order  of
perturbation    theory    for    the    operators   $\underline{U}_g$,
$\hat{\varphi}(x)$, $R(x|J)$, $\underline{W}[\overline{\Phi}]$ will be
defined   non-uniquely   due   to  choice  of  representation.

In qunatum   field   theory,   one   usually   uses   the   asymptotic
in-representation: one supposes that the particles become free at
$t\to - \infty$. Therefore, the state space can be identified with the
free Fock space of in-particles. For this case, the Heisenberg field
$\hat{\varphi}(x)$ weakly  tends  as $t\to\pm\infty$ to the asymptotic
free field:
\bez
\hat{\varphi}(x) \sim_{x^0  \to  -\infty} \hat{\varphi}_{in}(x) \equiv
\hat{\varphi}_0(x);
\quad
\hat{\varphi}(x) \sim_{x^0  \to  +\infty} \hat{\varphi}_{out}(x).
\eez
The $S$-matrix  (it will be denoted as
$\hat{\Sigma}_0$) is  an  unitary  transformation  between  asymptotic
fields:
\beq
\hat{\varphi}_{out}(x) =
\hat{\Sigma}_0^+    \hat{\varphi}_{in}(x) \hat{\Sigma}_0 =
\hat{\Sigma}_0^+    \hat{\varphi}_{0}(x) \hat{\Sigma}_0.
\l{5.1}
\eeq
Denote
\beq
\hat{\Sigma}_{\overline{\Phi}} \equiv                   \hat{\Sigma}_0
\underline{W}[\overline{\Phi}].
\l{5.2}
\eeq
Making use of these notations and assumptions, write down the formulas
for perturbation theory:
\beq
\underline{U}_{g_1g_2} = \underline{U}_{g_1} \underline{U}_{g_2},
\quad
\underline{U}_{g^{-1}} \hat{\varphi}_0(x)       \underline{U}_g      =
\hat{\varphi}_0 (w_gx);
\l{5.2a}
\eeq
\beb
\underline{U}_g \hat{\Sigma}_{\overline{\Phi}} \underline{U}_{g^{-1}} =
\hat{\Sigma}_{u_g\overline{\Phi}},
\quad
\hat{\Sigma}_{\overline{\Phi}}^+ =
\hat{\Sigma}_{\overline{\Phi}}^{-1};\\
\frac{\delta}{\delta \overline{\Phi}(y)}
\left(
\hat{\Sigma}_{\overline{\Phi}}^+
\frac{\delta \hat{\Sigma}_{\overline{\Phi}}^+}
{\delta \overline{\Phi}(x)}
\right) = 0, \qquad y \gsim x.
\l{5.2b}
\eeb
As   $J\sim     0$, the $R$-function
$R(x|J)    \equiv
R\{x|\overline{\Phi}\}$ should satisfy
the Yang-Feldman relation:
\beq
(\partial_{\mu} \partial^{\mu} + V'{}'(\overline{\Phi}(x)))
(R\{x|\overline{\Phi}\} - \overline{\Phi}(x)) =
-ih
\hat{\Sigma}_{\overline{\Phi}}^+
\frac{\delta \hat{\Sigma}_{\overline{\Phi}}^+}
{\delta \overline{\Phi}(x)}
\l{5.3}
\eeq
and boundary conditions:
\beb
R\{x|\overline{\Phi}\} \sim_{x^0      \to      -\infty}       \sqrt{h}
\hat{\varphi}_0(x);\\
R\{x|\overline{\Phi}\} \sim_{x^0      \to      +\infty}
\underline{W}^+[\overline{\Phi}]
\sqrt{h} \hat{\varphi}_{out}(x)
\underline{W}[\overline{\Phi}] =
\hat{\Sigma}_{\overline{\Phi}}^+
\sqrt{h} \hat{\varphi}_0(x)
\hat{\Sigma}_{\overline{\Phi}}.
\l{5.4}
\eeb
Therefore, the $R$-function can be presented in two forms,
which are corollaries of boundary conditions \r{5.4}:
\beb
R\{x|\overline{\Phi}\} - \overline{\Phi}(x) =
\hat{\phi}_-(x|\overline{\Phi}) \sqrt{h}
- ih \int dy
D^{ret}_{\overline{\Phi}}(x,y)
\hat{\Sigma}_{\overline{\Phi}}^+
\frac{\delta \hat{\Sigma}_{\overline{\Phi}}}
{\delta \overline{\Phi}(y)};
\\
R\{x|\overline{\Phi}\} - \overline{\Phi}(x) =
\hat{\Sigma}_{\overline{\Phi}}^+
\left[
\hat{\phi}_+(x|\overline{\Phi}) \sqrt{h}
- ih \int dy
D^{adv}_{\overline{\Phi}}(x,y)
\frac{\delta \hat{\Sigma}_{\overline{\Phi}}}
{\delta \overline{\Phi}(y)}
\hat{\Sigma}_{\overline{\Phi}}^+
\right]
\hat{\Sigma}_{\overline{\Phi}}.
\l{5.5}
\eeb
We obtain the following important identity:
\beq
\hat{\phi}_+(x|\overline{\Phi}) \hat{\Sigma}_{\overline{\Phi}} -
\hat{\Sigma}_{\overline{\Phi}} \hat{\phi}_-(x|\overline{\Phi})
= - i\sqrt{h} \int dy D_{\overline{\Phi}}(x,y)
\frac{\delta                    \hat{\Sigma}_{\overline{\Phi}}}{\delta
\overline{\Phi}(y)},
\l{5.6}
\eeq
since the commutation function has the form:
\bez
D_{\overline{\Phi}}(x,y) =
D^{ret}_{\overline{\Phi}}(x,y) -
D^{adv}_{\overline{\Phi}}(x,y).
\eez

{\bf 5.2.} Let $\hat{\varphi}_0(x)$ be a scalar free field of the mass
$m$,  $\underline{U}_g$ be the Poincare transformation in this theory,
Let  us show that the $S$-matrix properties \r{5.2b} and \r{5.6} imply
all other relations of section 3.

At $J\sim 0$,  define $R(x|J) \equiv R\{x|\overline{\Phi}\}$ from  the
first  formula  \r{5.5}.  Because of the Bogoliubov causality property
\r{5.2b},  the $R(x|J)$-function depends  on  $J(y)$  at  $y<x$  only.
Therefore,  the  definition can be extended to the case $J\not\sim 0$:
set $R(x|J)  \equiv  R(x|J+J_+)$,  where  $supp  J_+  \gsim  supp  J$,
$J+J_+\sim  0$.  The  causality  property \r{3.3b} remains valid.  The
Poincare invariance property \r{3.3a} is a corollary of covariance  of
relation \r{5.5}.

Set $\hat{\varphi}(x)  \sqrt{h}  \equiv  R(x|0)$  and  check  relation
\r{3.12e}. Rewrite it (at $x\gsim supp \overline{\Phi}$) as
\bez
\hat{\Sigma}_0 \hat{\varphi}(x) \sqrt{h} \hat{\Sigma}_0^+ =
\hat{\Sigma}_{\overline{\Phi}}
\left[
\hat{\phi}_-(x|\overline{\Phi}) \sqrt{h}
- ih \int dy
D^{ret}_{\overline{\Phi}}(x,y)
\hat{\Sigma}_{\overline{\Phi}}^+
\frac{\delta \hat{\Sigma}_{\overline{\Phi}}}
{\delta \overline{\Phi}(y)}
\right]
\hat{\Sigma}_{\overline{\Phi}}^+
\eez
or (because of \r{5.6})
\beq
\hat{\Sigma}_0 \hat{\varphi}(x) \sqrt{h} \hat{\Sigma}_0^+ =
\hat{\phi}_+(x|\overline{\Phi}) \sqrt{h}
- ih \int dy
D^{adv}_{\overline{\Phi}}(x,y)
\frac{\delta \hat{\Sigma}_{\overline{\Phi}}}
{\delta \overline{\Phi}(y)}
\hat{\Sigma}_{\overline{\Phi}}^+
\l{5.7}
\eeq
It follows from the causality relation that
$\frac{\delta \hat{\Sigma}_{\overline{\Phi}}}
{\delta \overline{\Phi}(y)}
\hat{\Sigma}_{\overline{\Phi}}^+$
depends only on
$\overline{\Phi}(z)$ for $z>y$.
Since $\overline{\Phi}(y) =  0$  at  $y>x$,
property \r{5.7} is taken to the form
\beq
\hat{\Sigma}_0 \hat{\varphi}(x) \sqrt{h} \hat{\Sigma}_0^+ =
\hat{\phi}_+(x|0) \sqrt{h}
- ih \int dy
D^{adv}_{0}(x,y)
\frac{\delta \hat{\Sigma}_{\overline{\Phi}}}
{\delta \overline{\Phi}(y)}|_{\overline{\Phi} = 0}
\hat{\Sigma}_{0}^+
\l{5.8}
\eeq
However, the  property  \r{5.8}  is a corollary of \r{5.5} and \r{5.6}
for $\overline{\Phi} = 0$.

The commutation relation
\r{3.3c}   is checked by a direct computation.
Its sketch is as follows.

1. One writes relation \r{3.3c} in the more convenient form:
\beb
[\int dx R(x|J) \delta_1J(x),\int dy R(y|J) \delta_2J(y)] = \\
-ih \int dx
[\delta_2R(x|J) \delta_1J(x) - \delta_1R(x|J) \delta_2J(x)];
\l{5.9}
\eeb
for $R(x|J)$, one uses the first relation \r{5.5}:
\beb
R(x|J) = \overline{\Phi}(x) + \sqrt{h} r_-(x|\overline{\Phi}); \\
r_-(x|\overline{\Phi}) =
\hat{\phi}_-(x|\overline{\Phi}) +
\sqrt{h} \int         dy        D^{ret}_{\overline{\Phi}}        (x,y)
\hat{j}(y|\overline{\Phi});\\
\hat{j}(y|\overline{\Phi}) = \frac{1}{i}
\hat{\Sigma}_{\overline{\Phi}}^+
\frac{\delta \hat{\Sigma}_{\overline{\Phi}}^+}
{\delta \overline{\Phi}(y)}.
\l{5.9a}
\eeb

2. One takes the right-hand side
of \r{5.9} to the form:
\beb
-ih \{
\int dx
[\delta_2\overline{\Phi}(x)        \delta_1J(x)       -
\delta_1\overline{\Phi}(x) \delta_2J(x)]
\\
- \sqrt{h} \int dxdz D^{ret}_{\overline{\Phi}}(x,z)
V'{}'{}'(\overline{\Phi}(z)) r_-(z|\overline{\Phi})
[\delta_2\overline{\Phi}(z)        \delta_1J(x)       -
\delta_1\overline{\Phi}(z) \delta_2J(x)]
\\
+ h \int dxdy D^{ret}_{\overline{\Phi}}(x,y)
[\delta_2\hat{j}(y|\overline{\Phi}) \delta_1J(x)
- \delta_1\hat{j}(y|\overline{\Phi}) \delta_2J(x)]
\}
\l{5.10}
\eeb

3. One variates the relation
\bez
\hat{\Sigma}_{\overline{\Phi}}^+
\hat{\phi}_+(x|\overline{\Phi})
\hat{\Sigma}_{\overline{\Phi}} -
\hat{\phi}_-(x|\overline{\Phi}) =
\sqrt{h} \int dy D_{\overline{\Phi}}(x,y)
\hat{j}(y|\overline{\Phi})
\eez
with respect to
$\overline{\Phi}(z)$. After all computations,
one obtains:
\beq
-i [\hat{j}(z|\overline{\Phi}),\hat{\phi}_-(x|\overline{\Phi})] =
\sqrt{h} \int dy D_{\overline{\Phi}}(x,y)
\frac{\delta \hat{j}(z|\overline{\Phi})}{\delta \overline{\Phi}(y)}
- D_{\overline{\Phi}}(x,y) V'{}'{}'(\overline{\Phi}(z))
\hat{r}_-(z|\overline{\Phi}).
\l{5.11}
\eeq
Further, it follows from unitarity that
\beq
[ \hat{j}(\xi|\overline{\Phi});
\hat{j}(z|\overline{\Phi}) ] =
- i
\left(
\frac{\delta \hat{j}(\xi|\overline{\Phi})}{\delta \overline{\Phi}(z)}
-
\frac{\delta \hat{j}(z|\overline{\Phi})}{\delta \overline{\Phi}(\xi)}
\right)
\l{5.12}
\eeq

4. Making use of the commutation relations
\r{5.11} and \r{5.12}, one takes the left-hand side
of eq.\r{5.9} to the form \r{5.10}.

{\bf 5.3.}    Relation \r{5.6}  can be viewed as a basis
of the semiclassical perturbation theory.
It is convenient to consider the substitution
\bez
\hat{\Sigma}_{\overline{\Phi}} =                    W[\overline{\Phi}]
\tilde{\Sigma}_{\overline{\Phi}};
\eez
then formula
\r{5.6} will be taken to the form
\beq
[\hat{\phi}_-(x|\overline{\Phi}) ; \tilde{\Sigma}_{\overline{\Phi}}] =
\sqrt{h} \int dy D_{\overline{\Phi}}(x,y) \left\{
\hat{\rho}(y|\overline{\Phi}) \tilde{\Sigma}_{\overline{\Phi}} - i
\frac{\delta                  \tilde{\Sigma}_{\overline{\Phi}}}{\delta
\overline{\Phi}(y)}
\right\},
\l{5.13}
\eeq
where $\hat{\rho}(y|\overline{\Phi})$   has the form
\r{4.20}.  Expand
$\tilde{\Sigma}_{\overline{\Phi}}$ into an asymptotic
series in $\sqrt{h}$:
\bez
\tilde{\Sigma}_{\overline{\Phi}} =        1         +         \sqrt{h}
\tilde{\Sigma}^{(1)}_{\overline{\Phi}} +
h \tilde{\Sigma}^{(2)}_{\overline{\Phi}} + ...
\eez
Therefore, for  the  $k$-th  order  of  the  perturbation theory,  one
obtains the following recursive relations:
\beq
[\hat{\phi}_-(x|\overline{\Phi}) ; \tilde{\Sigma}^{(k)}_{\overline{\Phi}}] =
\sqrt{h} \int dy D_{\overline{\Phi}}(x,y) \left\{
\hat{\rho}(y|\overline{\Phi}) \tilde{\Sigma}^{(k-1)}_{\overline{\Phi}}
- i \frac{\delta \tilde{\Sigma}^{(k-1)}_{\overline{\Phi}}}{\delta
\overline{\Phi}(y)}
\right\}.
\l{5.14}
\eeq

Investigate whether there exists
$\tilde{\Sigma}_{\overline{\Phi}}^{(k)}$ satisfying eq.\r{5.14}.
It is convenient to use symbolic calculus based on
the notion of a normal symbol \c{Z}.
One expands the operators into a series
containing normal products of the fields
$\hat{\phi}_-(x|{\overline{\Phi}})$:
\beq
\tilde{\Sigma}_{\overline{\Phi}}^{(k)} = \sum_m \int dx_1...dx_m
\tilde{\Sigma}_{\overline{\Phi},m}^{(k)} (x_1,...,x_m)
:\hat{\phi}_-(x_1|{\overline{\Phi}}) ...
\hat{\phi}_-(x_m|{\overline{\Phi}}):
\eeq
By $\tilde{S}^{(k)}_{\overline{\Phi}}[\phi_-(\cdot)]$
we denote the normal symbol:
\bez
\tilde{S}_{\overline{\Phi}}^{(k)}[\phi_-(\cdot)]
= \sum_m \int dx_1...dx_m
\tilde{\Sigma}_{\overline{\Phi},m}^{(k)} (x_1,...,x_m)
{\phi}_-(x_1) ... {\phi}_-(x_m).
\eez
One also has
\bez
\tilde{\Sigma}_{\overline{\Phi}}^{(k)} =
:\tilde{S}_{\overline{\Phi}}^{(k)}[\phi_-(\cdot|{\overline{\Phi}})]:
\eez

Introduce the following notations. By
$A*B$ we denote the normal symbol of the
propduct of operators
$\hat{A} = :A[\hat{\phi}_-(\cdot|{\overline{\Phi}})]:$
and $\hat{B} = :B[\hat{\phi}_-(\cdot|{\overline{\Phi}})]:$. It is
equal to
\beq
(A*B)[\phi_-(\cdot)] =
A[\hat{\phi}_-(\cdot) +          \frac{1}{i}          \int          dy
D_{\overline{\Phi}}^-(\cdot,y) \frac{\delta}{\delta        \phi_-(y)}]
B[\phi_-(\cdot)],
\l{5.16}
\eeq
by $\frac{DA}{D{\overline{\Phi}}(x)}$ let us denote
the normal symbol of the operator
$\frac{\delta    \hat{A}}{\delta     {\overline{\Phi}}(x)}$.
Its explicit form is
\beq
\frac{DA}{D{\overline{\Phi}}(x)} [\phi_-(\cdot)] =
\frac{\delta A}{\delta {\overline{\Phi}}(x)}[\phi_-(\cdot)]
- \int              d\xi              D^{ret}_{\overline{\Phi}}(\xi,x)
V'{}'{}'({\overline{\Phi}}) \phi_-(x)      \frac{\delta      A}{\delta
\phi_-(\xi)}[\phi_-(\cdot)]
\l{5.17a}
\eeq
Since the normal symbol of the commutator
$[\hat{\phi}_-(x|{\overline{\Phi}});\hat{A}]$ has the form
$\int dy \frac{1}{i} D_{\overline{\Phi}}(x,y)
\frac{\delta A}{\delta \phi_-(y)}$,
property \r{5.14} is taken to the form
\beq
\frac{\delta\tilde{S}^{(k)}_{\overline{\Phi}}}{\delta \phi_-(y)} =
i \rho(y|{\overline{\Phi}})  *  \tilde{S}^{(k-1)}_{\overline{\Phi}}  +
\frac{D\tilde{S}^{(k-1)}_{\overline{\Phi}}}{D{\overline{\Phi}}(y)},
\l{5.17}
\eeq
where
\bez
\rho(y|{\overline{\Phi}}) = \rho(y|{\overline{\Phi}},\phi_-) =
- \frac{1}{2}    V'{}'{}'({\overline{\Phi}}(y))     [\phi_-^2(y)     +
\alpha(y|{\overline{\Phi}})] -
\eez
is a normal symbol of the operator
$\hat{\rho}(y|{\overline{\Phi}})$.

Relation
\r{5.17} can be solved and
$\tilde{S}^{(k)}[\phi]$  can be found under the following condition
\beq
[
i\rho(y|{\overline{\Phi}}) * + \frac{D}{D{\overline{\Phi}}(y)};
i\rho(z|{\overline{\Phi}}) * + \frac{D}{D{\overline{\Phi}}(z)}
] = 0.
\l{5.18}
\eeq
It follows from \r{5.17a} that
$[\frac{D}{D{\overline{\Phi}}(y)}
;\frac{D}{D{\overline{\Phi}}(z)}] =  0$, so that
relation \r{5.18} is taken to the form:
\beq
i \left(
\frac{D\rho(z|{\overline{\Phi}})}{D{\overline{\Phi}}(y)}
-
\frac{D\rho(y|{\overline{\Phi}})}{D{\overline{\Phi}}(z)}
\right)
=
\rho(y|{\overline{\Phi}}) * \rho(z|{\overline{\Phi}})
- \rho(z|{\overline{\Phi}}) * \rho(y|{\overline{\Phi}}).
\l{5.19}
\eeq
Formula
\r{5.19} is another form of
eq.\r{4.29a}. It is checked in analogous way,
It implies that there exist
$\tilde{S}^{(k)}_{\overline{\Phi}}$  and
$\tilde{\Sigma}^{(k)}_{\overline{\Phi}}$ satisfying
eqs. \r{5.17} and \r{5.14}.

The operator
$\tilde{\Sigma}_{\overline{\Phi}}^{(k)}$  is found form
eq.\r{5.14} up to a multiplier by a c-number
$a^{(k)}_{\overline{\Phi}}$; therefore,
\beq
\tilde{\Sigma}_{\overline{\Phi}}^{(k)} =
\tilde{\Sigma}_{\overline{\Phi} 0}^{(k)} + a^{(k)}_{\overline{\Phi}},
\l{5.20}
\eeq
here $\tilde{\Sigma}^{(k)}_{{\overline{\Phi}}0}$
is uniquely determined from the normalization condition
\bez
<0|\tilde{\Sigma}^{(k)}|0> = 0.
\eez

Investigate now the properties of Poincare invariance,  unitarity  and
causality. Notice that the operator
$U_g
\tilde{\Sigma}^{(k)}_{u_g{\overline{\Phi}}} U_g^{-1}$
satisfies eq.
\r{5.14} and coincides up to an
additional c-number constant with
$\tilde{\Sigma}^{(k)}_{\overline{\Phi}}$.
Therefore, the property of Poincare invariance is satisfied if
\beq
a^{(k)}_{u_g{\overline{\Phi}}} = a^{(k)}_{\overline{\Phi}}.
\l{5.21}
\eeq
To check the unitarity property, notice that the operator
$\tilde{\Sigma}_{\overline{\Phi}}$ obeying relation
\r{5.13} also satisfies the condition
\beq
[\hat{\phi}_-(x|{\overline{\Phi}});
\tilde{\Sigma}_{\overline{\Phi}}^+
\tilde{\Sigma}_{\overline{\Phi}}] =
\sqrt{h} \int dy D_{\overline{\Phi}}(x,y) (-i)
\frac{\delta}{\delta {\overline{\Phi}}(y)}
(\tilde{\Sigma}_{\overline{\Phi}}^+
\tilde{\Sigma}_{\overline{\Phi}}).
\l{5.22}
\eeq
By
$(\tilde{\Sigma}_{\overline{\Phi}}^+
\tilde{\Sigma}_{\overline{\Phi}})_k$
we denote the
$k$-th order of perturbative expansion of the operator
$(\tilde{\Sigma}_{\overline{\Phi}}^+
\tilde{\Sigma}_{\overline{\Phi}})$
into a series in $\sqrt{h}$. Then the commutator
$[\hat{\phi}_-(x|{\overline{\Phi}});
(\tilde{\Sigma}_{\overline{\Phi}}^+
\tilde{\Sigma}_{\overline{\Phi}})_k]$
will be expressed from
\r{5.22} via
$(\tilde{\Sigma}_{\overline{\Phi}}^+
\tilde{\Sigma}_{\overline{\Phi}})_{k-1}$.
It will  vanish if the unitarity property is satisfied in the previous
order of the perturbation theory. Therefore, the operator
$(\tilde{\Sigma}_{\overline{\Phi}}^+
\tilde{\Sigma}_{\overline{\Phi}})_k$
is a multiplicator by a c-number. It vanishes if
\beq
0 =
<0|
(\tilde{\Sigma}_{\overline{\Phi}}^+
\tilde{\Sigma}_{\overline{\Phi}})_k|0> \equiv
2 Re a^{(k)}_{\overline{\Phi}} +
\sum_{s=1}^{k-1}
<0|
\tilde{\Sigma}_{\overline{\Phi}}^{(s)+}
\tilde{\Sigma}_{\overline{\Phi}}^{(k-s)}|0>.
\l{5.23}
\eeq
Relation
\r{5.23}  fixes the real part of
$a^{(k)}_{\overline{\Phi}}$ uniquely.

Investigate now the causality property. Making use of notations
\r{5.9a}, consider the commutation relation
\r{5.11}  at  $x\lsim  supp {\overline{\Phi}}$ и $x^0<z^0$:
\bez
[\hat{\varphi}_0(x);\hat{j}(z|{\overline{\Phi}})] =
- i\sqrt{h} \int dy D_{\overline{\Phi}}(x,y)
\frac{\delta \hat{j}(x|{\overline{\Phi}})}{\delta {\overline{\Phi}}(y)}
+ i D_{\overline{\Phi}}(x,z) V'{}'{}'({\overline{\Phi}}(z))
\hat{r}_-(z|{\overline{\Phi}}).
\eez
Therefore, for the variational derivative
$\frac{\delta                     \hat{j}(z|{\overline{\Phi}})}{\delta
{\overline{\Phi}}(\xi)}$ при $\xi^0>z^0$ one has:
\bez
[\varphi_0(x);
\frac{\delta                \hat{j}^{(k)}(z|{\overline{\Phi}})}{\delta
{\overline{\Phi}}(\xi)}] =
- i \int dy
\frac{\delta}{\delta {\overline{\Phi}}(\xi)}
\left(
D_{\overline{\Phi}}(x,y)
\frac{\delta              \hat{j}^{(k-1)}(z|{\overline{\Phi}})}{\delta
{\overline{\Phi}}(y)},
\right)
\eez
where $\hat{j}^{(k)}(z|{\overline{\Phi}})$ is the
$k$-th order of perturbation theory for
$\hat{j}(z|{\overline{\Phi}})$. Therefore,  the  Bogoliubov  causality
condition in the
$(k-1)$-th order implies that the $k$-th order for the operator
$\frac{\delta                \hat{j}^{(k)}(z|{\overline{\Phi}})}{\delta
{\overline{\Phi}}(\xi)}$ at   $z^0   <\xi^0$   will   be   a  c-number
multiplier. Therefore,  the causality property  is  satisfied  in  the
$k$-th order iff
\beq
\frac{\delta}{\delta {\overline{\Phi}}(\xi)}
<0|\hat{j}^{(k)}(z|{\overline{\Phi}})|0> = 0,
\quad \xi \gsim z.
\l{5.24}
\eeq
Let us calculate the vacuum average of the operator
$\hat{j}^{(k)}$. Present eq.\r{5.9a} as
\beq
\hat{j}(x|{\overline{\Phi}}) =      \tilde{\Sigma}^+_{\overline{\Phi}}
\hat{\rho}(x|{\overline{\Phi}}) \tilde{\Sigma}_{\overline{\Phi}} +
\frac{1}{i} \tilde{\Sigma}^+_{\overline{\Phi}}
\frac{\delta                  \tilde{\Sigma}_{\overline{\Phi}}}{\delta
{\overline{\Phi}}(x)}.
\l{5.25}
\eeq
Therefore,
\beq
\frac{1}{i} <0|
\frac{\delta                  \tilde{\Sigma}_{\overline{\Phi}}}{\delta
{\overline{\Phi}}(x)}|0> =
<0| \tilde{\Sigma}_{\overline{\Phi}} \hat{j}(x|{\overline{\Phi}})|0>
-                                   <0|\hat{\rho}(x|{\overline{\Phi}})
\tilde{\Sigma}_{\overline{\Phi}}|0>,
\l{5.26}
\eeq
and $\hat{\rho}(x|{\overline{\Phi}})                              =
\hat{j}^{(0)}(x|{\overline{\Phi}})$.
Consider the $k$-th order of the perturbation theory for
\r{5.26}. We come to the following relations:
\beb
<0|\hat{j}^{(k)}(x|{\overline{\Phi}})|0> =
\frac{1}{i} \frac{\delta a^{(k)}_{\overline{\Phi}}}
{\delta {\overline{\Phi}}(x)}
- \sum_{s=1}^{k-1} <0|
\tilde{\Sigma}^{(s)}_{\overline{\Phi}}
\hat{j}^{(k-s)}(x|{\overline{\Phi}})|0> -
<0|[\tilde{\Sigma}^{(k)}_{\overline{\Phi}};
\hat{\rho}(x|{\overline{\Phi}})]|0>.
\\ \qquad\l{5.27}
\eeb
Properties \r{5.23}  and  \r{5.27} allows us to
obtain the conditions on imaginary part of
$a^{(k)}_{\overline{\Phi}}$:
\beb
\frac{1}{i}
\frac{\delta^2 a^{(k)}_{\overline{\Phi}}}{\delta  {\overline{\Phi}}(x)
\delta {\overline{\Phi}}(y)} =
\frac{\delta}{\delta {\overline{\Phi}}(y)}
\left[
\sum_{s=1}^{k-1} <0|\tilde{\Sigma}^{(s)}_{\overline{\Phi}}
\hat{j}^{(k-s)}(x|{\overline{\Phi}})|0>
+
<0|[\tilde{\Sigma}^{(k)}_{\overline{\Phi}};
\hat{\rho}(x|{\overline{\Phi}})]|0>
\right], \quad y \gsim x.
\\ \qquad
\l{5.28}
\eeb

Thus, for each order of the perturbation theory one  should  construct
the   operator   $\tilde{\Sigma}^{(k)}_{\overline{\Phi}}$   from   eq.
\r{5.14}   (or   \r{5.17})   and   choose   a   c-number    multiplier
$a^{(k)}_{\overline{\Phi}}$  from  the  Poincare  invariance condition
\r{5.21}, unitarity property \r{5.23} and causality relation \r{5.28}.
Then      the      properties     \r{5.2b}     of     the     operator
$\hat{\Sigma}_{\overline{\Phi}}$  will   be   satisfied   within   the
perturbation framework.

The c-number functional $a^{(k)}_{\overline{\Phi}}$ is defined up to a
purely       imaginary       Poincare       invariant       functional
$i\gamma^{(k)}_{\overline{\Phi}}$  satisfying  the  locality  property
$\frac{\delta^2        \gamma^{(k)}_{\overline{\Phi}}}         {\delta
{\overline{\Phi}}(x) \delta {\overline{\Phi}}(y)} = 0$,  $x\ne y$. The
non-uniqueness  is  related   with   the   possibility   of   one-loop
renormalization of the Lagrangian,

{\bf 5.4}.  To  illustrate the obtained relations,  consider the lower
orders of perturbation theory.

The zero order gives the relation
$\tilde{S}^{(0)}_{\overline{\Phi}} =  1$;  the first order
for relation \r{5.17} implies that
\bez
\frac{\delta \tilde{S}^{(1)}_{\overline{\Phi}}}{\delta  \phi_-(y)} = -
\frac{i}{2} V'{}'{}'({\overline{\Phi}}(y))       [\phi_-^2(y)        +
\alpha(y|{\overline{\Phi}})];
\eez
therefore,
\bez
\tilde{S}^{(1)}_{\overline{\Phi}} =
- i \int dx
V'{}'{}'({\overline{\Phi}}(x))
\left[
\frac{1}{6} \phi_-^3(x)  +   \frac{1}{2}   \alpha(x|{\overline{\Phi}})
\phi_-(x)
\right] + a_{\overline{\Phi}}^{(1)}
\eez
The unitarity and causality properties
\r{5.23}  and   \r{5.28} will be written as
\bez
Re a_{\overline{\Phi}}^{(1)} = 0,
\quad
\frac{1}{i}
\frac{\delta^2 a_{{\overline{\Phi}}}^{(1)}}
{\delta {\overline{\Phi}}(x) \delta {\overline{\Phi}}(y)} =
\frac{\delta}{\delta {\overline{\Phi}}(y)}
<0|[
\tilde{\Sigma}^{(1)}_{\overline{\Phi}};
\hat{\rho}(x|{\overline{\Phi}})]|0>.
\eez
Therefore, it is possible to choose
$a_{\overline{\Phi}}^{(1)} = 0$.

The second  order  of  perturbation  theory  gives  us  the  following
relations
\bey
\rho(y|{\overline{\Phi}}) * \tilde{S}^{(1)}_{\overline{\Phi}} =
\\
\frac{i}{2} V'{}'{}'({\overline{\Phi}}(y))
(\phi_-^2(y) + \alpha(y|{\overline{\Phi}}))
\int dx
V'{}'{}'({\overline{\Phi}}(x))
\left[
\frac{1}{6} \phi_-^3(x)  +   \frac{1}{2}   \alpha(x|{\overline{\Phi}})
\phi_-(x)
\right]
\\
+ \frac{1}{2} V'{}'{}'({\overline{\Phi}}(y)) \phi_-(y)
\int dx V'{}'{}'({\overline{\Phi}}(x))
D^-_{\overline{\Phi}}(y,x)
[\phi_-^2(x) + \alpha(x|{\overline{\Phi}})]
\\
- \frac{i}{2} V'{}'{}'({\overline{\Phi}}(y))
\int dx [D^-_{\overline{\Phi}}(y,x)]^2
V'{}'{}'({\overline{\Phi}}(x)) \phi_-(x);
\eey
\bey
\frac{D\tilde{S}^{(1)}}{D{\overline{\Phi}}(y)} =
- i V^{(IV)}({\overline{\Phi}}(y))
\left[
\frac{1}{6} \phi_-^3(y)  +   \frac{1}{2}   \alpha(y|{\overline{\Phi}})
\phi_-(y)
\right] \\
- \frac{i}{2} \int dx V'{}'{}'({\overline{\Phi}}(x))
\phi_-(x) \frac{\delta            \alpha(x|v)}{\delta            v(y)}
V'{}'{}'({\overline{\Phi}}(y))
\\
- \int               dx               D^{ret}_{{\overline{\Phi}}}(x,y)
V'{}'{}'({\overline{\Phi}}(y)) \phi_-(y)
(-\frac{i}{2} V'{}'{}'({\overline{\Phi}}(x)))
[\phi_-^2(x) + \alpha(x|{\overline{\Phi}})].
\eey
Combining the results, we obtain from eq.
\r{5.17}   the following realtion for
$\tilde{S}^{(2)}$:
\bey
\frac{\delta \tilde{S}^{(2)}}{\delta \phi_-(y)} =
- i V^{(IV)}({\overline{\Phi}}(y))
\left[
\frac{1}{6} \phi_-^3(y)  +   \frac{1}{2}   \alpha(y|{\overline{\Phi}})
\phi_-(y)
\right] \\
- \frac{1}{2} V'{}'{}'({\overline{\Phi}}(y))
(\phi_-^2(y) + \alpha(y|{\overline{\Phi}}))
\int dx V'{}'{}'({\overline{\Phi}}(x))
\left[
\frac{1}{6} \phi_-^3(x)  +   \frac{1}{2}   \alpha(x|{\overline{\Phi}})
\phi_-(x)
\right] \\
+ \frac{i}{2} V'{}'{}'({\overline{\Phi}}(y))
\phi_-(y) \int dx V'{}'{}'({\overline{\Phi}}(x))
[D^-_{\overline{\Phi}} + D^{adv}_{\overline{\Phi}}](y,x)
(\phi_-^2(x) + \alpha(x|{\overline{\Phi}}))
\\
+ \frac{1}{2} V'{}'{}'({\overline{\Phi}}(y))
\int dx V'{}'{}'({\overline{\Phi}}(x)) \phi_-(x)
\left\{
[D_{\overline{\Phi}}^-(y,x)]^2 +
\frac{1}{i} \frac{\delta \alpha\{x|v\}}{\delta v(y)}
\right\}
\eey
For the simplicity, introduce the notations:
\beb
D^c_{\overline{\Phi}} \equiv          D^-_{\overline{\Phi}}          +
D^{adv}_{\overline{\Phi}}; \\
(D^c_{\overline{\Phi}} (y,x))^2_R \equiv
[D^-_{\overline{\Phi}}(y,x)]^2 +
\frac{1}{i} \frac{\delta \alpha\{x|v\}}{\delta v(y)}.
\l{5b.1}
\eeb
Formally, it follows from
eq.\r{4.29} that
\beb
D^c_{\overline{\Phi}}(y,x) =
\theta(x^0-y^0) D^-_{\overline{\Phi}}(x,y)
\theta(y^0-x^0) D^-_{\overline{\Phi}}(y,x);\\
(D^c_{\overline{\Phi}}(y,x))^2_R =
\theta(x^0-y^0) (D^-_{\overline{\Phi}}(x,y))^2
\theta(y^0-x^0) (D^-_{\overline{\Phi}}(y,x))^2 =
(D^c_{\overline{\Phi}}(y,x))^2;
\l{5b.2}
\eeb
However, formulas \r{5b.2} should be clarified at $x=y$ due to divergences.

One has:
\bey
\tilde{S}^{(2)} = - i \int dx V^{(IV)}({\overline{\Phi}}(x))
\left[
\frac{1}{24} \phi_-^4(x)  +  \frac{1}{4}   \alpha(x|{\overline{\Phi}})
\phi_-^2 \right] \\
- \frac{1}{4} \left\{
\int dx V'{}'{}'({\overline{\Phi}}(x))
\left[
\frac{1}{6} \phi_-^3(x)  +   \frac{1}{2}   \alpha(x |{\overline{\Phi}})
\phi_-(x)
\right]
\right\}^2
\\
+ \frac{i}{4} \int dxdy
V'{}'{}'({\overline{\Phi}}(x))
V'{}'{}'({\overline{\Phi}}(y)) D^c_{\overline{\Phi}}(x,y)
\\ \times \left(
[\phi_-^2(x) + \alpha(x|{\overline{\Phi}})]
[\phi_-^2(y) + \alpha(y|{\overline{\Phi}})] -
\alpha(x|{\overline{\Phi}}) \alpha(y|{\overline{\Phi}})
\right)
\\
+ \frac{1}{2} \int dx dy
V'{}'{}'({\overline{\Phi}}(x))
V'{}'{}'({\overline{\Phi}}(y)) (D^c_{\overline{\Phi}}(x,y))^2_R
\phi_-(x) \phi_-(y) + a^{(2)}_{\overline{\Phi}}.
\eey

The c-number
$a^{(2)}_{\overline{\Phi}}$
satisfies the  properties  of  Poincare  invariance,   unitarity   and
causality. It happens that one can look for
$a^{(2)}_{\overline{\Phi}}$ in the following form:
\bey
a^{(2)}_{\overline{\Phi}} =       -      \frac{i}{8}      \int      dx
V^{(IV)}({\overline{\Phi}}(x)) (\alpha(x|{\overline{\Phi}}))^2
\\
- \frac{1}{8} \int dxdy
V'{}'{}'({\overline{\Phi}}(x))
V'{}'{}'({\overline{\Phi}}(y))
\alpha(x|{\overline{\Phi}}) \alpha(y|{\overline{\Phi}})
\frac{1}{i} D^c_{\overline{\Phi}}(x,y)
\\
- \frac{1}{12} \int dxdy
V'{}'{}'({\overline{\Phi}}(x))
V'{}'{}'({\overline{\Phi}}(y))
(\frac{1}{i} D^c_{\overline{\Phi}}(x,y))^3_R
\eey
where $(D_{\overline{\Phi}}^c)^3_R$ is a
renormalized cub of the function
$D^c_{\overline{\Phi}}$.

The unitarity condition
\r{5.23} can be written as
\bez
Re a^{(2)}_{\overline{\Phi}}         =          -          \frac{1}{2}
<0|\tilde{\Sigma}^{(1) +}_{\overline{\Phi}}
\tilde{\Sigma}^{(1)}_{\overline{\Phi}}|0>;
\eez
it leads to the relation
\beq
Re (\frac{1}{i} D^c_{\overline{\Phi}}(x,y))^3_R =
\frac{1}{2}
\left[
(\frac{1}{i} D^-_{\overline{\Phi}}(x,y))^3
+ (\frac{1}{i} D^-_{\overline{\Phi}}(y,x))^3
\right],
\l{5b.3}
\eeq
and
\beq
Re (\frac{1}{i} D^c_{\overline{\Phi}}(x,y)) =
\frac{1}{2}
\left[
(\frac{1}{i} D^-_{\overline{\Phi}}(x,y))
+ (\frac{1}{i} D^-_{\overline{\Phi}}(y,x))
\right],
\l{5b.4}
\eeq
Relation \r{5b.4}  is  obviously  satisfied;  property  \r{5b.3} is an
important condition on $(D^c_{\overline{\Phi}})^3_R$.

Consider the vacuum average value of
$\hat{j}^{(2)}$  according to formula \r{5.27}:
\bez
<0|\hat{j}^{(2)}(\xi|{\overline{\Phi}})|0> =
\frac{1}{i} \frac{\delta             a^{(2)}_{\overline{\Phi}}}{\delta
{\overline{\Phi}}(\xi)}
-                            <0|\tilde{\Sigma}^{(1)}_{\overline{\Phi}}
\hat{j}^{(1)}(\xi|{\overline{\Phi}})|0> -
<0|[\tilde{\Sigma}^{(2)}_{\overline{\Phi}};
\hat{\rho}(\xi|{\overline{\Phi}})]|0>.
\eez
The normal symbol
\bez
\hat{j}^{(1)}(\xi|{\overline{\Phi}}) =
\frac{1}{i} \frac{\delta \tilde{\Sigma}^{(1)}_{\overline{\Phi}}}
{\delta {\overline{\Phi}}(\xi)} +
[\hat{\rho}(\xi);\tilde{\Sigma}^{(1)}_{\overline{\Phi}}]
\eez
has the form
\bey
j^{(1)}(\xi|{\overline{\Phi}},\phi_-) =
- V^{(IV)} ({\overline{\Phi}}(\xi))
\left[
\frac{1}{6} \phi_-^3(\xi) + \frac{1}{2}  \alpha(\xi|{\overline{\Phi}})
\phi_-(\xi)
\right]
\\
+ \frac{1}{2} V'{}'{}'({\overline{\Phi}}(\xi)) \phi_-(\xi)
\int dx V'{}'{}'({\overline{\Phi}}(x))
D^{ret}_{\overline{\Phi}}(\xi,x)
[\phi_-^2(x) + \alpha(x|{\overline{\Phi}})]
\\
- \frac{1}{2} V'{}'{}'({\overline{\Phi}}(\xi))
\int dx V'{}'{}'({\overline{\Phi}}(x)) \phi_-(x)
\frac{\delta \alpha\{\xi|v\}}{\delta v(x)}.
\eey
Therefore,
\bey
<0|\hat{j}^{(2)}(\xi|{\overline{\Phi}})|0> =
- \frac{1}{8}                          V^{(V)}({\overline{\Phi}}(\xi))
(\alpha(\xi|{\overline{\Phi}})^2
- \frac{1}{4}       \int       dx       V^{(IV)}({\overline{\Phi}}(x))
V'{}'{}'({\overline{\Phi}}(\xi)) \alpha(x|{\overline{\Phi}})
\frac{\delta \alpha\{\xi|v\}}{\delta v(x)}
\\
- \frac{1}{6}      \int      dy       V^{(IV)}({\overline{\Phi}}(\xi))
V'{}'{}'({\overline{\Phi}}(y))
[(D^c_{\overline{\Phi}}(\xi,y))^3_R   -
(D^-_{\overline{\Phi}}(y,\xi))^3]
\\
+ \frac{1}{4} \int dy
V^{(IV)}({\overline{\Phi}}(\xi))
V'{}'{}'({\overline{\Phi}}(y)) \alpha\{\xi|v\} \alpha\{y|v\}
[(D^c_{\overline{\Phi}}(\xi,y))   -
(D^-_{\overline{\Phi}}(y,\xi))]
\\
+ \frac{1}{4} \int dx dy
V'{}'{}'({\overline{\Phi}}(x))
V'{}'{}'({\overline{\Phi}}(\xi))
V'{}'{}'({\overline{\Phi}}(y))
G^1_v(\xi,x,y)
\\
+ \frac{1}{4} \int dx dy
V'{}'{}'({\overline{\Phi}}(x))
V'{}'{}'({\overline{\Phi}}(\xi))
V'{}'{}'({\overline{\Phi}}(y))
\alpha\{x|v\} \frac{\delta \alpha\{\xi|v\}}{\delta v(y)}
[D^c_{\overline{\Phi}}(x,y) - D^-_{\overline{\Phi}}(x,y)]
\\
+ \frac{1}{8} \int dxdy
V'{}'{}'({\overline{\Phi}}(x))
V'{}'{}'({\overline{\Phi}}(\xi))
V'{}'{}'({\overline{\Phi}}(y))
\alpha\{x|v\} \alpha\{y|v\} G^2_v(\xi,x,y).
\eey
Here
\bey
G^1_v(\xi,x,y) =
- \frac{1}{3} \frac{\delta}{\delta v(\xi)}
(D^c_{\overline{\Phi}}(x,y))^3_R -
D^{ret}_{\overline{\Phi}}(\xi,y) D^-_{\overline{\Phi}}(x,\xi)
(D^-_{\overline{\Phi}}(x,y))^2 \\
- D^{ret}_{\overline{\Phi}}(\xi,x) D^-_{\overline{\Phi}}(y,\xi)
(D^-_{\overline{\Phi}}(y,x))^2 -
(D^c_{\overline{\Phi}}(x,y))^2_R
(D^-_{\overline{\Phi}}(y,\xi) D^-_{\overline{\Phi}}(x,\xi)
- D^-_{\overline{\Phi}}(\xi,y) D^-_{\overline{\Phi}}(\xi,x)),
\eey
\bey
G^2(\xi,x,y) =
\frac{\delta D^c_{\overline{\Phi}}(x,y)}{\delta v(\xi)}
+ D^{ret}_{\overline{\Phi}}(\xi,y) D^-_{\overline{\Phi}}(x,\xi)
+ D^{ret}_{\overline{\Phi}}(\xi,x) D^-_{\overline{\Phi}}(y,\xi)
\\
+ D^-_{\overline{\Phi}}(x,\xi) D^-_{\overline{\Phi}}(y,\xi)
- D^-_{\overline{\Phi}}(\xi,x) D^-_{\overline{\Phi}}(\xi,y) =
- D^{ret}_{\overline{\Phi}}(\xi,x) D^{ret}_{\overline{\Phi}}(\xi,y).
\eey
The causality condition \r{5.24} is satisfied, provided that
\beb
G^1(\xi,x,y) = 0, \mbox{ при } \xi\lsim x \mbox{ или } \xi\lsim y;\\
(D^c_{\overline{\Phi}}(\xi,y))^3_R =
(D^c_{\overline{\Phi}}(\xi,y))^3 \mbox{ при } \xi\lsim y.
\l{5b.6}
\eeb
Notice that formally
$(D^c_{\overline{\Phi}}(\xi,y))^3_R =
(D^c_{\overline{\Phi}}(\xi,y))^3$; moreover,
\bey
G^1(\xi,x,y) =                        D^{ret}_{\overline{\Phi}}(\xi,x)
D^{ret}_{\overline{\Phi}}(\xi,y) (D^c_{\overline{\Phi}}(x,y))^2_R
\\
- D^{ret}_{\overline{\Phi}}(\xi,y) D^-_{\overline{\Phi}}(x,\xi)
[(D^-_{\overline{\Phi}}(x,y))^2 -
(D^c_{\overline{\Phi}}(x,y))^2]
\\
- D^{ret}_{\overline{\Phi}}(\xi,x) D^-_{\overline{\Phi}}(y,\xi)
[(D^-_{\overline{\Phi}}(y,x))^2 -
(D^c_{\overline{\Phi}}(x,y))^2]
\eey
and conditions \r{5b.6} are satisifed. However, the function
$(D^c_{\overline{\Phi}})^3_R$ contains the divergences
which are eliminated by the renormalization procedure,
so that relations
\r{5b.4}   and  \r{5b.6}  are important for renormalization.

\section{Unstable particles}

In the previous section,  we have supposed that the usual  assumptions
of the  $S$-matrix theory are satisfied.  On the other hand,  when one
investigate the bound states and unstable  particles,  the  $S$-matrix
conception leads  to  difficulties.  Let  us  show  that semiclassical
perturbation theory can be developed even for  the  unstable  particle
case. Consider a simple example. Let the action of the model be
\bez
I[\Phi,X] = \int dx
\left[
\frac{1}{2} \partial_{\mu} \Phi \partial^{\mu} \Phi +
\frac{1}{2} \partial_{\mu} X \partial^{\mu} X -
\frac{m^2}{2}\Phi^2 - \frac{M^2}{2} X^2 + \Phi^2 X
\right].
\eez
By $\hat{\varphi}(x)$,  $\hat{\chi}(x)$ we  denote  the  corresponding
quantum  fields.  It  is  well-known  that  for  the case $M>2m$,  one
$X$-particle can decay into  two  $\Phi$-particles.  Investigate  this
process.

By $a_{\bf p}^{\pm}$ we denote the creation and annihilation operators
for the free $\Phi$-particles. Let
$b^{\pm}_{\bf   p}$  be creation and annihilation operators for
the free $X$-particles.
Then the zeroth order of perturbation theory gives the relation
\bey
\hat{\chi}_0(x) =
\frac{1}{(2\pi)^{d/2}} \int \frac{d{\bf p}}{\sqrt{2\Omega_{\bf p}}}
[b^+_{\bf p} e^{i\Omega_{\bf p} t - i{\bf p}x} +
b^-_{\bf p} e^{-i\Omega_{\bf p} t + i{\bf p}x}]; \quad
\Omega_{\bf p} = \sqrt{{\bf p}^2 + M^2};\\
\hat{\varphi}_0(x) =
\frac{1}{(2\pi)^{d/2}} \int \frac{d{\bf p}}{\sqrt{2\omega_{\bf p}}}
[a^+_{\bf p} e^{i\omega_{\bf p} t - i{\bf p}x} +
a^-_{\bf p} e^{-i\omega_{\bf p} t + i{\bf p}x}]; \quad
\omega_{\bf p} = \sqrt{{\bf p}^2 + m^2}.
\eey

Investigate the operator of the form $\hat{\chi}(x)$ in the firt order
of perturbation theory:
\bez
\hat{\chi} = \hat{\chi}_0 + \sqrt{h}\hat{\chi}_1 + ...
\eez
For zero claical  fields,  the  Yang-Feldman  relation  \r{3.12d}  for
the quantum field
$\hat{\chi}(x)$  takes the form:
\bez
(\partial_{\mu}\partial^{\mu} + M^2) \chi(x) =
- i\sqrt{h}
\frac{\delta \underline{W}[\overline{\Phi},\overline{X}]}
{\delta \overline{X}(x)}|_{\overline{\Phi},\overline{X}=0};
\eez
therefore,
\beq
(\partial_{\mu}\partial^{\mu} + M^2) \chi_1(x) =
- i
\frac{\delta {W}[\overline{\Phi},\overline{X}]}
{\delta \overline{X}(x)}|_{\overline{\Phi},\overline{X}=0}.
\l{6.3}
\eeq
Analogously to  eq.\r{4.20},  we  obtain  that  the right-hand side of
relation \r{6.3} has the form
\bey
-i W^+
\frac{\delta W}{\delta \overline{X}(x)} =
\\
:\left[
\frac{1}{2} \frac{\partial^3V}{\partial \Phi^2 \partial X}
\hat{\phi}_-^2(x|\overline{\Phi},\overline{X})
+ \frac{\partial^3V}{\partial \Phi \partial X^2}
\hat{\phi}_-(x|\overline{\Phi},\overline{X})
\hat{\chi}_-(x|\overline{\Phi},\overline{X}) +
\frac{1}{2} \frac{\partial^3V}{\partial X^3}
\hat{\chi}_-^2(x|\overline{\Phi},\overline{X})
\right]:
+ \alpha(x|\overline{\Phi},\overline{X})
\eey
As $\overline{\Phi},\overline{X} = 0$,
\bez
- i \frac{\delta W}{\delta \overline{X}(x)}
|_{\overline{\Phi},\overline{X} =0}  =  -   :\hat{\varphi}_0^2(x):   +
\alpha_0(x).
\eez
In particular, the positive-frequency part
$\hat{\chi}_1^{++}(x)$ with two creation operators
$a^+$ obeys the following equation:
\beb
(\partial_{\mu} \partial^{\mu} + M^2) \hat{\chi}_1^{++}(x) =
- (\hat{\varphi}_0(x))^2 =
\\
- \frac{1}{(2\pi)^d} \int
\frac{d{\bf k}_1}{\sqrt{2\omega_{{\bf k}_1}}}
\frac{d{\bf k}_2}{\sqrt{2\omega_{{\bf k}_2}}}
a^+_{{\bf k}_1}
a^+_{{\bf k}_2}
e^{i(\omega_{{\bf k}_1} + \omega_{{\bf k}_2})t - i({\bf  k}_1  +  {\bf
k}_2){\bf x}}.
\l{6.4}
\eeb
The solution of eq.\r{6.4} is non-unique,  since we have not fixed the
representation of  canonical  commutation relations.  Let us fix a
representation. First, require that the momentum operator has the form:
\bez
{\bf P} = \int d{\bf k} {\bf k} (a^+_{\bf k}a^-_{\bf k} + b^+_{\bf  k}
b^-_{\bf k}),
\eez
Then the operator $\hat{\chi}_1^{++}$ will have the form:
\bey
\hat{\chi}_1^{++}(x) =
- \frac{1}{(2\pi)^d} \int
\frac{d{\bf k}_1}{\sqrt{2\omega_{{\bf k}_1}}}
\frac{d{\bf k}_2}{\sqrt{2\omega_{{\bf k}_2}}}
a^+_{{\bf k}_1}
a^+_{{\bf k}_2}
\\ \times
\left[
\frac{1}{\Omega_{{\bf k}_1 +{\bf k}_2}^2 -
(\omega_{{\bf k}_1} + \omega_{{\bf k}_2})^2}
e^{i(\omega_{{\bf k}_1} + \omega_{{\bf k}_2})t - i({\bf  k}_1  +  {\bf
k}_2){\bf x}}
+ \tilde{g}_{{\bf k}_1 {\bf k}_2}
e^{i\Omega_{{\bf k}_1 + {\bf k}_2}t - i({\bf  k}_1  +  {\bf
k}_2){\bf x}}
\right]
\eey
where $\tilde{g}_{{\bf  k}_1{\bf  k}_2}$
is an  arbitrary function.  It depends on the particular choice of the
representation. The integrand should not contain the poles.
Therefore,
\bez
\tilde{g}_{{\bf k}_1 {\bf k}_2} =
- \frac{1}{\Omega_{{\bf k}_1 +{\bf k}_2}^2 -
(\omega_{{\bf k}_1} + \omega_{{\bf k}_2})^2}
+ g_{{\bf k}_1 {\bf k}_2},
\eez
where the function
$g_{{\bf k}_1{\bf  k}_2}$ has no poles.
Therefore,
\beb
\hat{\chi}_1^{++}(x) =
- \frac{1}{(2\pi)^d}
\int
\frac{d{\bf k}_1}{\sqrt{2\omega_{{\bf k}_1}}}
\frac{d{\bf k}_2}{\sqrt{2\omega_{{\bf k}_2}}}
a^+_{{\bf k}_1}
a^+_{{\bf k}_2}
e^{i\Omega_{{\bf k}_1 + {\bf k}_2}t - i({\bf  k}_1  +  {\bf
k}_2){\bf x}}
\\ \times
\left[
\frac{1}{\Omega_{{\bf k}_1 +{\bf k}_2}^2 -
(\omega_{{\bf k}_1} + \omega_{{\bf k}_2})^2}
(e^{i(\omega_{{\bf k}_1} + \omega_{{\bf k}_2} -
\Omega_{{\bf k}_1 + {\bf k}_2})t} - 1)
+ {g}_{{\bf k}_1 {\bf k}_2}
\right]
\l{6.5}
\eeb

Let us calculate now the decay rate..Let the  initial  condition  have
the form:
\beq
\Psi_0 \equiv
\int d{\bf x} dt \hat{\chi}({\bf x},t) \alpha({\bf x},t)|0>
\l{6.6}
\eeq
In the zeroth order of perturbation theory, one has
\bez
\Psi_0 \simeq \int d{\bf p}
\frac{\tilde{\alpha}({\bf p},\Omega_{{\bf p}})}
{\sqrt{2\Omega_{{\bf p}}}}
b^+_{\bf p}|0>,
\eez
with
\bez
\tilde{\alpha}({\bf p},{\varepsilon}) \equiv
\frac{1}{(2\pi)^{d/2}} \int d{\bf x}d\tau
e^{i{\varepsilon}\tau - i{\bf p}{\bf x}},
\eez
Therefore, the state contains one
$X$-particle with the wave function
\beq
\psi_0({\bf p}) =
\frac{\tilde{\alpha}({\bf p},\Omega_{{\bf p}})}
{\sqrt{2\Omega_{{\bf p}}}}.
\l{6.6a}
\eeq
Evaluate the probability amplitude that at time $t$ there will be two
$\Phi$-particles with momenta ${\bf k}_1$ и ${\bf k}_2$:
\beb
\psi_{{\bf k}_1{\bf k}_2}(t) =
\frac{1}{\sqrt{2}}
<0|a^-_{{\bf k}_1} a^-_{{\bf k}_2} e^{-iHt}|\Psi_0>
= \frac{1}{\sqrt{2}}
<0|a^-_{{\bf k}_1} a^-_{{\bf k}_2}
\int d{\bf   x}   d\tau   \hat{\chi}({\bf   x},\tau   -  t)\alpha({\bf
x},\tau)|0>; \\ \qquad
\l{6.7}
\eeb
Here the property $e^{-iHt} \hat{\chi}({\bf x},\tau) e^{iHt}
= \hat{\chi}({\bf   x},\tau-t)$   is   used.   The   only   nontrivial
contribution to the matrix element \r{6.7} is given by the  components
$\hat{\chi}_1^{++}$; therefore, it follows from eq.\r{6.5} that:
\bey
\psi_{{\bf k}_1{\bf k}_2} =
- \frac{\sqrt{2h}}{(2\pi)^{d/2}}
\frac{1}{\sqrt{2\omega_{{\bf k}_1}}}
\frac{1}{\sqrt{2\omega_{{\bf k}_2}}}
e^{-i(\omega_{{\bf k}_1} + \omega_{{\bf k}_2})t}
\\ \times
\left[
\frac{1}{\Omega_{{\bf k}_1+{\bf   k}_2}^2   -  (\omega_{{\bf  k}_1}  +
\omega_{{\bf k}_2})^2}
\tilde{\alpha}
({\bf k}_1 + {\bf k}_2, \omega_{{\bf k}_1} + \omega_{{\bf k}_2}) +
\right.
\\
\left.
\tilde{\alpha}
({\bf k}_1 + {\bf k}_2, \Omega_{{\bf k}_1 + {\bf k}_2})
e^{i(\omega_{{\bf k}_1} + \omega_{{\bf k}_2} -
\Omega_{{\bf k}_1 + {\bf k}_2}) t}
\left(
g_{{\bf k}_1{\bf k}_2} -
\frac{1}{\Omega_{{\bf k}_1+{\bf   k}_2}^2   -  (\omega_{{\bf  k}_1}  +
\omega_{{\bf k}_2})^2}
\right)
\right]
\eey
Consider the rate of transition
\bez
\Gamma_{{\bf k}_1{\bf  k}_2}  =   \frac{1}{t}   |\psi_{{\bf   k}_1{\bf
k}_2}(t)|^2
\eez
Let $t\to\infty$. Making use of formula
$\left|
\frac{e^{i\alpha t} - 1}{\alpha}
\right|^2 \frac{1}{t} \simeq 2\pi \delta(\alpha)$,
one finds that
\bez
\Gamma_{{\bf k}_1{\bf k}_2} \simeq
\frac{2h}{(2\pi)^d}
\frac{1}{{2\omega_{{\bf k}_1}}}
\frac{1}{{2\omega_{{\bf k}_2}}}
\frac{1}{{(2\Omega_{{\bf k}_1 +{\bf k}_2}})^2}
2\pi
\delta(\omega_{{\bf k}_1} + \omega_{{\bf k}_2} -
\Omega_{{\bf k}_1 + {\bf k}_2})
|\tilde{\alpha}
({\bf k}_1 + {\bf k}_2, \Omega_{{\bf k}_1 + {\bf k}_2})|^2.
\eez
Substitute the initial wave function:
\bez
\Gamma_{{\bf k}_1{\bf k}_2} \simeq
\frac{2h}{(2\pi)^d}
\frac{1}{{2\omega_{{\bf k}_1}}}
\frac{1}{{2\omega_{{\bf k}_2}}}
\frac{1}{{2\Omega_{{\bf k}_1 +{\bf k}_2}}}
2\pi
\delta(\omega_{{\bf k}_1} + \omega_{{\bf k}_2} -
\Omega_{{\bf k}_1 + {\bf k}_2})
|\psi_0({\bf k}_1 + {\bf k}_2)|^2
\eez
Let the initial momentum of the particle be equal to
${\bf P}$; then
$|\psi_0({\bf p})|^2   =   \delta({\bf  p}  -  {\bf  P})$,
and the standard formula for the decay rate is reproduced.

Thus, the  semiclassical  approach  allows us to reproduce the quantum
field theory results for decay rates.

\section{Conclusions}

Thus, semiclassical  perturbation  field  theory  can  be  constructed
analogously to the axiomatic  quantum  field  theory.  The  axioms  of
Poincare covariance, unitarity and Bogoliubov causality are formulated
analogously; however, the correspondence principle between quantum and
classical theories gives new relations. Therefore, the scattering
matrix on the external background field is calculated in each order of
perturbation expansion  up  to  {\it a  c-number}  local  term,  not  up  to
quasilocal {\it operator.}

The semiclassical perturbation theory may be generalized to  the  case
of unstable particles.

The generalization  to Fermi fields is not difficult:  one should only
substitute commuting variables by Grassmannian variables in a standard
way. The  case of constrained systems and gauge fields is non-trivial;
the author is going to discuss it in further publications.

\newpage
\pagestyle{empty}

\end{document}